\documentclass{emulateapj}
\usepackage[modulo]{lineno}
\usepackage[xcolor]{xcolor}
\usepackage{amsmath}

\newcommand{\iso}[2]{\hbox{${}^{#1}{\rm #2}$}}

\def\hd{\object[HD 94028]{HD~94028}}
\def\hda{\object[HD 108317]{HD~108317}}

\def\hde{\object[HD 196944]{HD~196944}}
\def\hdf{\object[HD 201626]{HD~201626}}
\def\hdonesix{\object[HD 160617]{HD~160617}}

\def\kmsec{\mbox{km~s$^{\rm -1}$}}
\def\logg{\mbox{log~{\it g}}}
\def\msun{\mbox{$M_{\odot}$}}
\def\teff{\mbox{$T_{\rm eff}$}}
\def\vt{\mbox{$v_{\rm t}$}}
\def\cemps{\mbox{CEMP-$s$}}
\def\rpro{\mbox{$r$-process}}
\def\spro{\mbox{$s$-process}}
\def\ipro{\mbox{$i$-process}}
\def\ncap{\mbox{$n$-capture}}
\def\loggf{\mbox{$\log gf$}}

\shorttitle{Rare Elements in HD 94028}
\shortauthors{Roederer et al.}

\begin{document}

\title{
The Diverse Origins of Neutron-Capture Elements
in the Metal-Poor Star HD 94028:\ \\
Possible Detection of Products of $i$-process 
Nucleosynthesis\footnotemark[$\dagger$]
}

\footnotetext[$\dagger$]{Some
data presented in this paper were obtained from the 
Barbara A.\ Mikulski Archive for Space Telescopes (MAST).~
The Space Telescope Science Institute 
is operated by the Association of Universities for Research in 
Astronomy, Inc., under NASA contract NAS5-26555. 
These data are associated with Programs GO-7402 and GO-8197.
Based on data obtained from the European Southern Observatory (ESO) 
Science Archive Facility.
These data are associated with Program 072.B-0585(A).
This paper includes data taken at The McDonald Observatory of The 
University of Texas at Austin.
}

\author{
Ian U.\ Roederer,\altaffilmark{1,2}
Amanda I.\ Karakas,\altaffilmark{3,4} 
Marco Pignatari,\altaffilmark{5,6,7}
Falk Herwig\altaffilmark{2,7,8}
}

\altaffiltext{1}{Department of Astronomy, University of Michigan,
1085 S.\ University Ave., Ann Arbor, MI 48109, USA;
\mbox{iur@umich.edu}
}
\altaffiltext{2}{Joint Institute for Nuclear Astrophysics and Center for the
Evolution of the Elements (JINA-CEE), USA
}
\altaffiltext{3}{Research School of Astronomy \& Astrophysics, 
the Australian National University, Canberra ACT 2611, Australia
}
\altaffiltext{4}{Monash Centre for Astrophysics, 
School of Physics and Astronomy, Monash University, VIC 3800, Australia;
\mbox{amanda.karakas@monash.edu}
}
\altaffiltext{5}{E.A.\ Milne Centre for Astrophysics, 
Department of Physics \& Mathematics, University of Hull, 
HU6 7RX, United Kingdom;
\mbox{mpignatari@gmail.com}
}
\altaffiltext{6}{Konkoly Observatory, Research Centre for Astronomy 
and Earth Sciences, Hungarian Academy of Sciences, 
Konkoly Thege Miklos ut 15-17, H-1121 Budapest, Hungary;
}
\altaffiltext{7}{NuGrid collaboration, \url{http://www.nugridstars.org}
}
\altaffiltext{8}{Department of Physics \& Astronomy, 
University of Victoria, Victoria, BC, V8P5C2 Canada;
\mbox{fherwig@uvic.ca}
}


\addtocounter{footnote}{9}

\begin{abstract}

We present a detailed analysis of the 
composition and nucleosynthetic origins 
of the heavy elements in the metal-poor
([Fe/H]~$= -$1.62~$\pm$~0.09) star HD~94028.
Previous studies revealed that this star
is mildly enhanced in elements produced by the 
slow neutron-capture process ($s$~process; 
e.g., [Pb/Fe]~$= +$0.79~$\pm$~0.32)
and
rapid neutron-capture process ($r$~process;
e.g., [Eu/Fe]~$= +$0.22~$\pm$~0.12), 
including unusually large molybdenum ([Mo/Fe]~$=+$0.97~$\pm$~0.16)
and ruthenium ([Ru/Fe]~$=+$0.69~$\pm$~0.17) enhancements.
However, this star is not enhanced in carbon
([C/Fe]~$= -$0.06~$\pm$~0.19).
We analyze an archival near-ultraviolet spectrum of HD~94028,
collected using the 
Space Telescope Imaging Spectrograph on board the
\textit{Hubble Space Telescope},
and other archival
optical spectra collected from ground-based telescopes.
We report abundances or upper limits
derived from 64~species of 56~elements.
We compare these observations with 
$s$-process yields from low-metallicity 
AGB evolution and nucleosynthesis models.
No combination of $s$- and $r$-process patterns
can adequately reproduce the observed abundances,
including the super-solar [As/Ge] ratio ($+$0.99~$\pm$~0.23)
and the enhanced [Mo/Fe] and [Ru/Fe] ratios.
We can fit these features 
when including an additional contribution 
from the intermediate neutron-capture process ($i$~process),
which perhaps operated 
by the ingestion of 
H in He-burning convective regions
in massive stars, super-AGB stars,
or low-mass AGB stars.
Currently, only the $i$~process appears capable of 
consistently producing the super-solar [As/Ge] 
ratios and ratios among neighboring heavy elements
found in HD~94028.
Other metal-poor stars also show enhanced [As/Ge] ratios,
hinting that operation of the $i$~process may have been
common in the early Galaxy.

\end{abstract}

\keywords{
nuclear reactions, nucleosynthesis, abundances ---
stars: abundances ---
stars: AGB and post-AGB stars ---
stars: individual (HD~94028)
}

\section{Introduction}
\label{intro}

Understanding the origin of the elements remains
one of the major challenges of modern astrophysics. 
Ideally, to study the origin of each element,
it is desirable to collect data on each element's
abundance beyond the solar system.
Nearly all previous abundance studies of
heavy elements in the near-ultraviolet (NUV)
spectra of late-type stars
have focused on stars enriched 
by products of rapid neutron-capture (\rpro) nucleosynthesis
\citep{cowan96,cowan02,cowan05,sneden98,sneden03,
roederer09,roederer10a,roederer12a,roederer12d,roederer14d,roederer12b,
barbuy11,peterson11,siqueiramello13}.
One study \citep{placco14}
has studied the NUV spectrum of
a carbon-enhanced metal-poor (CEMP)
star with no enhancement of neutron-capture elements,
and another \citep{placco15}
studied two CEMP stars
enhanced with slow neutron-capture (\spro) material 
(\cemps\ stars).
Both stars in 
the latter
study, \hde\ and \hdf,
are C-enhanced ([C/Fe]~$ = +$1.1 and $+$1.5)
and show substantial enhancements in many elements
produced by \spro\ nucleosynthesis
(e.g., [Ba/Fe]~$= +$1.2 and $+$1.7; 
[Ce/Fe]~$= +$1.1 and $+$1.9;
[Pb/Fe]~$= +$2.1 and $+$2.9).

The new advance enabled by access to the NUV spectra of these two stars
was the detection of Ge~\textsc{i},
Nb~\textsc{ii}, Mo~\textsc{ii}, Cd~\textsc{i},
Lu~\textsc{ii}, Pt~\textsc{i}, and Au~\textsc{i}.
Most of these species are impossible to detect in the
spectral range accessible
to ground-based telescopes,
and Ge, Cd, Lu, Pt, and Au
had not been detected previously in any 
\cemps\ star.
\hde\ and \hdf\ are each in wide binary systems
with orbital periods of 1325 and 407~d 
\citep{lucatello05,placco15}.
\citeauthor{placco15}\ compared these abundances 
with model predictions, revealing
that each star could be reasonably fit
by the predicted \spro\ yields from models of low-mass
($\approx$~0.9~\msun) 
stars on the asymptotic giant branch (AGB;
\citealt{lugaro12,placco13}).~
The elements detected in the NUV 
had only a minor impact on the
model selection, but the abundances of
these elements 
all lie reasonably close (within $\approx$~2$\sigma$) 
to the model predictions.
One exception, Cd in \hde,
was about 1~dex lower than that predicted by the model.
Overall, however, the good agreement
was encouraging,
since the models had not previously
been confronted with observations of metal-poor stars
for these elements.

High-quality NUV spectra of stars capable
of studying the chemical fossil record 
are limited.
These data can only be collected with the
Goddard High Resolution Spectrograph (GHRS),
Space Telescope Imaging Spectrograph (STIS),
or Cosmic Origins Spectrograph (COS).~
STIS and COS are both currently in operation
on board the \textit{Hubble Space Telescope} (\textit{HST}),
and the GHRS was retired during Servicing Mission 2
in 1997.
The long exposure times required and stiff competition
for observing time on \textit{HST} 
present formidable challenges
to obtaining new observational data.
Fortunately, however, 
many high-quality spectroscopic datasets
exist in the \textit{HST} archives.
Some were taken for other scientific purposes,
and few abundance studies have focused specifically 
on the heaviest elements---those produced
mainly by \ncap\ reactions.

\hd\ is one such star with high-quality NUV spectra
in the \textit{HST} archives.
The high proper motion and $U-B$ excess of \hd\
have been known for decades
(e.g., \citealt{roman54}).
\citet{carney94} and \citet{clementini99} found
hints of radial velocity variations in their 
data for \hd,
but these were not confirmed by \citet{stryker85},
\citet{jasniewicz88}, \citet{latham02}, or \citet{roederer14c}.
\citet{peterson11} was the first to point out the
unusually-high Mo abundance of \hd,
[Mo/Fe]~$= +$1.0.
This star also appeared in the metal-poor sample
of \citet{roederer14c},
who found a mild excess of elements commonly produced 
by \spro\ nucleosynthesis.
Several investigators have extended the chemical
analysis of this star into the NUV,
but all have focused narrowly on one or a few elements:\
Be and B \citep{thorburn96,primas99};
Mo and Ru \citep{peterson11};
Ge, As, and Se \citep{roederer12c}; and
P \citep{jacobson14,roederer14f}.

In the present study, we perform a more thorough
analysis of all trace, heavy elements
detected in the NUV spectrum of \hd.
We place these results and the large set of 
optical abundances available from \citet{roederer14c}
on a single abundance scale.
The heavy elements in \hd\ show evidence of
enrichment by the $r$~process, $s$~process, and
possibly a third, still poorly-studied
nucleosynthesis process, the intermediate \ncap\ process ($i$~process).
We attempt to model this abundance pattern and
identify viable metal enrichment scenarios for \hd.

\section{Data}

\subsection{Observations from the Archives}
\label{observations}

We use two NUV spectroscopic data sets of \hd\ 
available in the 
Mikulski Archive for Space Telescopes.
These observations were made using 
STIS \citep{kimble98,woodgate98}
on board the \textit{HST}.~
One spectrum 
(data sets O5CN01-03, GO-8197, PI.\ Duncan)
has very high spectral resolution 
($R \equiv \lambda/\Delta\lambda \sim$~110,000).
This spectrum covers $\approx$~1885--2147~\AA\
with signal-to-noise (S/N) ratios 
$\approx$~35/1 per pixel near 2140~\AA.~
The other spectrum
(data sets O56D06-07, GO-7402, PI.\ Peterson)
has high spectral resolution
($R \sim$~30,000).
This spectrum covers 2280--3117~\AA\
with S/N ratios ranging from 
$\approx$~20 near 2300~\AA\ to
$\approx$~40 near 3100~\AA.~  
We use the reduction and coaddition provided by the
StarCAT database \citep{ayres10}.

\citet{roederer14c} derived abundances 
from an optical spectrum of \hd\
taken using the Robert G.\ Tull Coud\'{e} Spectrograph
on the Harlan J.\ Smith Telescope at McDonald Observatory, Texas.
We rederive abundances from this spectrum (Section~\ref{optical})
using the model atmosphere (Section~\ref{atm})
adopted in the present study.
This spectrum covers 3650--8000~\AA\
at $R \sim$~30,000
with S/N ratios ranging from
$\approx$~55 near 3950~\AA\ to
$\approx$~170 near 6750~\AA.~

We also use an optical spectrum taken with the
Ultraviolet and Visual Echelle Spectrograph (UVES; \citealt{dekker00}) 
on the Very Large Telescope (VLT) Kueyen at Cerro Paranal, Chile.
We obtained this spectrum from the ESO Science Archive.
This spectrum covers 3050--3860~\AA\
at $R \sim$~37,000
with S/N ratios ranging from
$\approx$~40 near 3200~\AA\ to
$\approx$~130 near 3800~\AA.~

\subsection{Atomic Data}
\label{atomic}

We compile our list of lines to examine from several recent 
studies of the NUV spectra of metal-poor stars.
These lines and the list of references are given in
Table~\ref{linetab}.
Many of these lines represent resonance or low-lying levels,
and they are frequently the dominant decay channels
from excited upper levels.
Thus, the uncertainty in the transition probability
is generally small and limited by the measurement uncertainty of
the lifetime of the upper level.
These uncertainties are also listed in Table~\ref{linetab}.

\begin{deluxetable}{cccccc}
\tablecaption{Abundances Derived from Lines Examined in This Study
\label{linetab}}
\tablewidth{0pt}
\tabletypesize{\scriptsize}
\tablehead{
\colhead{Species} &
\colhead{$\lambda$} &
\colhead{E.P.} &
\colhead{\loggf} &
\colhead{Ref.} &
\colhead{$\log \epsilon$} \\
\colhead{} &
\colhead{(\AA)} &
\colhead{(eV)} &
\colhead{} &
\colhead{} &
\colhead{} 
}
\startdata
Cu~\textsc{ii} & 2037.13 & 2.83 & $-$0.23  (0.03)  &  1 &  $+$2.25  (0.05)   \\
Cu~\textsc{ii} & 2054.98 & 2.93 & $-$0.29  (0.03)  &  1 &  $+$2.29  (0.15)   \\
Cu~\textsc{ii} & 2112.10 & 3.25 & $-$0.11          &  1 &  $+$2.60  (0.20)   \\
Cu~\textsc{ii} & 2126.04 & 2.83 & $-$0.23          &  1 &  $+$2.45  (0.10)   \\
As~\textsc{i}  & 1890.43 & 0.00 & $-$0.19          &  2 &  $+$1.10  (0.25)   \\
Mo~\textsc{ii} & 2015.11 & 0.00 & $-$0.49  (0.03)  &  3 &  $+$0.98  (0.15)   \\
Mo~\textsc{ii} & 2020.31 & 0.00 & $+$0.02  (0.04)  &  3 &  $+$1.32  (0.15)   \\
Mo~\textsc{ii} & 2045.97 & 0.00 & $-$0.35  (0.03)  &  3 &  $+$1.38  (0.15)   \\
Ru~\textsc{i}  & 3498.94 & 0.00 & $+$0.31  (0.02)  &  4 &  $+$0.69  (0.15)   \\
Pd~\textsc{i}  & 3404.58 & 0.81 & $+$0.33  (0.02)  &  5 &  $-$0.01  (0.20)   \\
Ag~\textsc{i}  & 3380.68 & 0.00 & $-$0.02  (0.01)  &  6 & $<-$0.35           \\
Ag~\textsc{i}  & 3382.90 & 0.00 & $-$0.33  (0.01)  &  6 & $<-$0.40           \\
Cd~\textsc{i}  & 2288.02 & 0.00 & $+$0.152 (0.013) &  7 &  $+$0.53  (0.20)   \\
Cd~\textsc{ii} & 2144.39 & 0.00 & $+$0.018 (0.002) &  1 &  $+$0.07  (0.20)   \\
Te~\textsc{i}  & 2142.82 & 0.00 & $-$0.32  (0.08)  &  8 &  $+$0.44  (0.30)   \\
Yb~\textsc{ii} & 2116.68 & 0.00 & $-$1.34          &  9 &  $-$0.26  (0.10)   \\
Yb~\textsc{ii} & 2126.74 & 0.00 & $-$0.87  (0.02)  & 10 &  $-$0.62  (0.15)   \\
Lu~\textsc{ii} & 2615.41 & 0.00 & $+$0.11  (0.04)  & 11 &  $-$0.50  (0.20)   \\
W~\textsc{ii}  & 2088.20 & 0.39 & $-$0.02  (0.02)  & 12 & $<-$0.40           \\
W~\textsc{ii}  & 2118.88 & 0.00 & $-$0.77  (0.04)  & 12 & $<-$0.05           \\
Os~\textsc{ii} & 2067.23 & 0.45 & $-$0.45  (0.03)  & 13 &  $+$0.14  (0.20)   \\
Os~\textsc{ii} & 2282.28 & 0.00 & $-$0.14  (0.03)  & 13 &  $+$0.12  (0.20)   \\
Pt~\textsc{i}  & 2067.51 & 0.00 & $-$0.62  (0.03)  & 14 &  $+$0.24  (0.15)   \\
Hg~\textsc{ii} & 1942.27 & 0.00 & $-$0.40  (0.04)  &  1 & $<+$0.30           \\
Pb~\textsc{i}  & 2833.05 & 0.00 & $-$0.50  (0.02)  & 15 &  $+$1.08  (0.30) 
\enddata
\tablerefs{
 (1) \citealt{roederer12b};
 (2) \citealt{holmgren75}, using HFS from \citealt{roederer12b};
 (3) \citealt{sikstrom01};
 (4) \citealt{wickliffe94};
 (5) \citealt{xu06};
 (6) \citealt{hansen12} for both \loggf\ value and HFS;
 (7) \citealt{morton00};
 (8) \citealt{roederer12a};
 (9) \loggf\ value from the Database on Rare Earths At Mons University
     (DREAM), using HFS/IS from \citealt{roederer12b};
(10) \citealt{kedzierski10}, using HFS from \citealt{roederer12b};
(11) \citealt{roederer10a}, using HFS from \citealt{roederer12d};
(12) \citealt{nilsson08};
(13) \citealt{ivarsson04};
(14) \citealt{denhartog05};
(15) \citealt{biemont00}, using HFS from \citealt{roederer12d}.
}
\end{deluxetable}

In Section~\ref{fewave}, we examine whether
any abundance trends with wavelength are apparent
in Fe~\textsc{i} or \textsc{ii} lines.
These lines are listed in Table~\ref{fetab}.
All lines used are included in the 
Atomic Spectra Database \citep{kramida14} maintained by the
National Institute of Standards and Technology (NIST).~
These lines have \loggf\ values rated with a
``C'' accuracy or better ($<$~25\%, or $<$~0.12~dex).
Table~\ref{fetab} also reports our
measurements of the equivalent widths of these lines.

\begin{deluxetable}{cccccc}
\tablecaption{NUV Iron Lines Examined
\label{fetab}}
\tablewidth{0pt}
\tabletypesize{\scriptsize}
\tablehead{
\colhead{Species} &
\colhead{$\lambda$} &
\colhead{E.P.} &
\colhead{\loggf} &
\colhead{Equiv.\ Width} &
\colhead{$\log \epsilon$} \\
\colhead{} &
\colhead{(\AA)} &
\colhead{(eV)} &
\colhead{} &
\colhead{(m\AA)} &
\colhead{} 
}
\startdata
  Fe~\textsc{i} & 2145.19 & 0.05 & $-$1.56 &  71.4 & 5.17  \\
  Fe~\textsc{i} & 2294.41 & 0.11 & $-$1.54 &  92.6 & 5.47  \\
  Fe~\textsc{i} & 2309.00 & 0.11 & $-$1.39 & 131.6 & 5.69  \\
  Fe~\textsc{i} & 2369.46 & 0.11 & $-$2.19 &  53.0 & 5.26
\enddata
\tablecomments{
The complete version of 
Table~\ref{fetab} is available in the online edition of the journal.
A portion is shown here to illustrate its form and content.
}
\end{deluxetable}

\section{Methods}

\subsection{Model Atmospheres}
\label{atm}

We adopt the model atmosphere parameters for \hd\
derived by \citet{roederer12c}, with
effective temperature \teff~$=$~5720~$\pm$~71~K,
surface gravity \logg~$=$~4.31~$\pm$~0.16,
microturbulence velocity \vt~$=$~0.90~$\pm$~0.30~\kmsec, and
metallicity [M/H]~$= -$1.62~$\pm$~0.09.
The uncertainties are estimated following
section~8.5 of \citet{roederer14c}.
The parallax of \hd\ is known to better than 5\%
(measured by \textit{Hipparcos}; \citealt{vanleeuwen07}),
and the inferred \logg\ value 
indicates \hd\ is an unevolved main sequence star.
Fe ionization equilibrium was not enforced
when deriving the model atmosphere parameters.
We interpolate the model atmosphere
from the ATLAS9 $\alpha$-enhanced grid of \citet{castelli03}.

\subsection{Analysis of NUV Data}
\label{uv}

We use a recent version of MOOG \citep{sneden73},
which includes the contribution of Rayleigh scattering from
atomic H in the source function \citep{sobeck11},
to derive the abundances.
For all lines in the NUV (except Fe~\textsc{i} and \textsc{ii} lines)
we use MOOG to generate a series of synthetic spectra
where the abundance of the element of interest
is adjusted to match the observed spectrum.
Our line lists for the syntheses 
begin with the \citet{kurucz95} lists,
and we update wavelengths and \loggf\ values
with more recent experimental values whenever possible.
Numerous unidentified transitions are found 
in the NUV spectra of late-type stars, like \hd.
Efforts to identify these transitions 
are underway \citep{peterson15},
and we include the results of these efforts in our line lists.
We model the remaining 
unidentified features using low-excitation Fe~\textsc{i} lines.

We include the effects of hyperfine splitting (HFS)
structure and isotope shifts (IS) in the
syntheses whenever possible.
We adopt isotopic mixtures that reflect
a predominantly \spro\ origin, using the
isotopic distributions presented in \citet{sneden08}.
For Cd~\textsc{ii}, the abundance derived using
the \spro\ isotopic mix differs from that using
the \rpro\ mix by $+$0.14~dex.
For Pb~\textsc{i}, the abundance derived using
the \spro\ isotopic mix differs from that using
the \rpro\ mix by $+$0.06~dex.
For both Yb~\textsc{ii} lines,
the \spro\ and \rpro\ isotopic mixtures
yield identical abundances.

\subsection{Re-analysis of Optical and NUV Data}
\label{optical}

\citet{roederer14c} performed a detailed abundance analysis
of \hd\ using the optical spectrum taken at 
McDonald Observatory.
The model atmosphere parameters adopted
by \citeauthor{roederer14c}\ are sufficiently different 
(\teff~$=$~5730~K, 
\logg~$=$~3.70,
\vt~$=$~1.00~\kmsec,
[M/H]~$= -$1.81)
that we have rederived 
abundances from all lines examined in that study
using the model atmosphere described in Section~\ref{atm}.
If \citeauthor{roederer14c}\ reported an equivalent width for a line,
we use MOOG to compute a theoretical equivalent width that 
is forced to match the measured equivalent width
by adjusting the input abundance.
For all other lines, we use the spectrum synthesis matching
technique described in Section~\ref{uv}.
We also rederive the P abundance from the P~\textsc{i} line at 2136~\AA,
and the result agrees 
with the value derived by \citet{roederer14f}.
The results of this reanalysis are
reflected in the abundances presented in Table~\ref{abundtab}.

\begin{deluxetable}{ccccccccc}
\tablecaption{Derived Abundances in HD~94028
\label{abundtab}}
\tablewidth{0pt}
\tabletypesize{\scriptsize}
\tablehead{
\colhead{Spec.} &
\colhead{N} &
\colhead{$\log \epsilon$} &
\colhead{[X/Fe]} &
\colhead{$\sigma_{\rm stat}$} &
\colhead{$\sigma_{\rm tot}$} &
\colhead{$\sigma_{\rm I}$} &
\colhead{$\sigma_{\rm II}$} &
\colhead{$\log \epsilon_{\odot}$} 
}
\startdata
 Fe~\textsc{i}  &  96 &     5.75 & $-$1.75 &  0.06 &  0.10 &  0.00 &  0.00 & 7.50 \\ 
 Fe~\textsc{ii} &  10 &     5.88 & $-$1.62 &  0.07 &  0.09 &  0.00 &  0.00 & 7.50 \\ 
 Li~\textsc{i}  &   1 &     2.03 & \nodata &  0.05 &  0.09 &  0.08 &  0.11 & ...  \\ 
 C\,(CH)        &   1 &     6.62 & $-$0.06 &  0.15 &  0.25 &  0.19 &  0.19 & 8.43 \\ 
 N\,(CN)        &   1 & $<$ 7.01 &$<$ 0.93 &\nodata&\nodata&\nodata&\nodata& 7.83 \\ 
 O~\textsc{i}   &   2 &     7.56 &    0.62 &  0.05 &  0.20 &  0.08 &  0.20 & 8.69 \\ 
 Na~\textsc{i}  &   2 &     4.57 &    0.08 &  0.12 &  0.22 &  0.14 &  0.23 & 6.24 \\ 
 Mg~\textsc{i}  &   4 &     6.19 &    0.34 &  0.05 &  0.24 &  0.09 &  0.22 & 7.60 \\ 
 Al~\textsc{i}  &   1 &     4.78 &    0.08 &  0.13 &  0.34 &  0.19 &  0.37 & 6.45 \\ 
 Si~\textsc{i}  &   2 &     6.15 &    0.39 &  0.18 &  0.26 &  0.19 &  0.26 & 7.51 \\ 
 P~\textsc{i}   &   1 &     4.05 &    0.39 &  0.16 &  0.19 &  0.18 &  0.18 & 5.41 \\ 
 K~\textsc{i}   &   1 &     3.61 &    0.33 &  0.12 &  0.23 &  0.14 &  0.23 & 5.03 \\ 
 Ca~\textsc{i}  &  10 &     4.94 &    0.35 &  0.10 &  0.25 &  0.14 &  0.25 & 6.34 \\ 
 Sc~\textsc{ii} &   6 &     1.71 &    0.18 &  0.05 &  0.13 &  0.20 &  0.09 & 3.15 \\ 
 Ti~\textsc{i}  &  14 &     3.40 &    0.20 &  0.05 &  0.20 &  0.08 &  0.20 & 4.95 \\ 
 Ti~\textsc{ii} &  22 &     3.74 &    0.41 &  0.05 &  0.14 &  0.20 &  0.09 & 4.95 \\ 
 V~\textsc{i}   &   1 &     2.21 &    0.03 &  0.11 &  0.22 &  0.13 &  0.22 & 3.93 \\ 
 V~\textsc{ii}  &   2 &     2.57 &    0.26 &  0.19 &  0.23 &  0.28 &  0.20 & 3.93 \\ 
 Cr~\textsc{i}  &  11 &     3.72 & $-$0.17 &  0.06 &  0.20 &  0.09 &  0.20 & 5.64 \\ 
 Cr~\textsc{ii} &   3 &     4.38 &    0.36 &  0.06 &  0.14 &  0.21 &  0.09 & 5.64 \\ 
 Mn~\textsc{i}  &   7 &     3.30 & $-$0.38 &  0.06 &  0.21 &  0.09 &  0.21 & 5.43 \\ 
 Co~\textsc{i}  &   2 &     3.12 & $-$0.12 &  0.12 &  0.24 &  0.14 &  0.24 & 4.99 \\ 
 Ni~\textsc{i}  &   6 &     4.52 &    0.05 &  0.12 &  0.23 &  0.14 &  0.23 & 6.22 \\ 
 Cu~\textsc{i}  &   1 &     1.96 & $-$0.48 &  0.13 &  0.15 &  0.15 &  0.24 & 4.19 \\ 
 Cu~\textsc{ii} &   4 &     2.30 & $-$0.27 &  0.10 &  0.14 &  0.13 &  0.12 & 4.19 \\ 
 Zn~\textsc{i}  &   3 &     3.08 &    0.27 &  0.07 &  0.20 &  0.10 &  0.21 & 4.56 \\ 
 Zn~\textsc{ii} &   1 &     3.26 &    0.32 &  0.10 &  0.11 &  0.11 &  0.11 & 4.56 \\ 
 Ge~\textsc{i}  &   3 &     1.56 & $-$0.34 &  0.17 &  0.20 &  0.19 &  0.20 & 3.65 \\ 
 As~\textsc{i}  &   2 &     1.18 &    0.63 &  0.13 &  0.16 &  0.15 &  0.16 & 2.30 \\ 
 Se~\textsc{i}  &   3 &     2.06 &    0.47 &  0.17 &  0.20 &  0.19 &  0.20 & 3.34 \\ 
 Rb~\textsc{i}  &   1 & $<$ 2.70 &$<$ 1.80 &\nodata&\nodata&\nodata&\nodata& 2.52 \\
 Sr~\textsc{ii} &   2 &     1.36 &    0.11 &  0.03 &  0.15 &  0.13 &  0.18 & 2.87 \\ 
 Y~\textsc{ii}  &   3 &     0.68 &    0.09 &  0.09 &  0.15 &  0.21 &  0.11 & 2.21 \\ 
 Zr~\textsc{ii} &   3 &     1.53 &    0.57 &  0.05 &  0.14 &  0.20 &  0.09 & 2.58 \\ 
 Nb~\textsc{ii} &   1 & $<$ 0.94 &$<$ 1.10 &\nodata&\nodata&\nodata&\nodata& 1.46 \\ 
 Mo~\textsc{i}  &   1 &     0.68 &    0.55 &  0.20 &  0.22 &  0.22 &  0.22 & 1.88 \\ 
 Mo~\textsc{ii} &   3 &     1.23 &    0.97 &  0.13 &  0.20 &  0.18 &  0.16 & 1.88 \\ 
 Tc~\textsc{i}  &   1 & $<$ 0.96 & \nodata &\nodata&\nodata&\nodata&\nodata& ...  \\ 
 Ru~\textsc{i}  &   1 &     0.69 &    0.69 &  0.15 &  0.18 &  0.17 &  0.18 & 1.75 \\ 
 Pd~\textsc{i}  &   1 &  $-$0.01 &    0.09 &  0.20 &  0.22 &  0.22 &  0.22 & 1.65 \\ 
 Ag~\textsc{i}  &   1 & $<-$0.40 & $<$0.15 &\nodata&\nodata&\nodata&\nodata& 1.20 \\ 
 Cd~\textsc{i}  &   1 &     0.53 &    0.57 &  0.20 &  0.25 &  0.22 &  0.24 & 1.71 \\ 
 Cd~\textsc{ii} &   1 &     0.07 & $-$0.02 &  0.20 &  0.25 &  0.24 &  0.22 & 1.71 \\ 
 Te~\textsc{i}  &   1 &     0.44 &    0.01 &  0.31 &  0.33 &  0.33 &  0.33 & 2.18 \\ 
 Ba~\textsc{ii} &   3 &     0.88 &    0.32 &  0.05 &  0.13 &  0.19 &  0.09 & 2.18 \\ 
 La~\textsc{ii} &   4 &  $-$0.28 &    0.24 &  0.12 &  0.18 &  0.24 &  0.14 & 1.10 \\ 
 Ce~\textsc{ii} &   2 &     0.25 &    0.29 &  0.13 &  0.19 &  0.24 &  0.16 & 1.58 \\ 
 Pr~\textsc{ii} &   1 & $<$ 0.35 &$<$ 1.25 &\nodata&\nodata&\nodata&\nodata& 0.72 \\ 
 Nd~\textsc{ii} &   1 &  $-$0.12 &    0.08 &  0.14 &  0.19 &  0.25 &  0.16 & 1.42 \\ 
 Sm~\textsc{ii} &   2 & $<-$0.35 &$<$ 0.31 &\nodata&\nodata&\nodata&\nodata& 0.96 \\ 
 Eu~\textsc{ii} &   2 &  $-$0.88 &    0.22 &  0.09 &  0.16 &  0.22 &  0.12 & 0.52 \\ 
 Gd~\textsc{ii} &   2 & $<-$0.10 &$<$ 0.45 &\nodata&\nodata&\nodata&\nodata& 1.07 \\ 
 Dy~\textsc{ii} &   1 &  $-$0.39 &    0.13 &  0.09 &  0.16 &  0.22 &  0.12 & 1.10 \\ 
 Er~\textsc{ii} &   2 & $<-$0.05 &$<$ 0.65 &\nodata&\nodata&\nodata&\nodata& 0.92 \\ 
 Tm~\textsc{ii} &   2 & $<-$0.95 &$<$ 0.57 &\nodata&\nodata&\nodata&\nodata& 0.10 \\ 
 Yb~\textsc{ii} &   3 &  $-$0.46 &    0.24 &  0.19 &  0.20 &  0.20 &  0.20 & 0.92 \\ 
 Lu~\textsc{ii} &   1 &  $-$0.62 &    0.90 &  0.20 &  0.25 &  0.24 &  0.22 & 0.10 \\ 
 Hf~\textsc{ii} &   2 & $<$ 0.15 &$<$ 0.92 &\nodata&\nodata&\nodata&\nodata& 0.85 \\ 
 W~\textsc{ii}  &   2 & $<-$0.40 &$<$ 0.57 &\nodata&\nodata&\nodata&\nodata& 0.65 \\ 
 Os~\textsc{ii} &   2 &     0.13 &    0.35 &  0.14 &  0.15 &  0.15 &  0.15 & 1.40 \\ 
 Ir~\textsc{i}  &   1 & $<$ 1.50 &$<$ 1.87 &\nodata&\nodata&\nodata&\nodata& 1.38 \\ 
 Pt~\textsc{i}  &   2 &     0.22 &    0.35 &  0.11 &  0.15 &  0.13 &  0.14 & 1.62 \\ 
 Hg~\textsc{ii} &   1 & $<$ 0.30 &$<$ 0.75 &\nodata&\nodata&\nodata&\nodata& 1.17 \\ 
 Pb~\textsc{i}  &   1 &     1.08 &    0.79 &  0.30 &  0.32 &  0.32 &  0.32 & 2.04    
\enddata

\end{deluxetable}

We use the UVES spectrum 
to derive abundances of additional elements
for the first time in \hd.
We derive these abundances using the spectrum synthesis method.
These lines are listed in Table~\ref{linetab}.

\citet{roederer14c} searched for lines that
yielded abundances consistently deviant from the mean 
abundance of each species.
That study defined a set of empirical corrections
for these 
optical lines,
listed in their Table~16,
and we apply the corrections
for metal-poor main sequence stars
to our rederived optical abundances when available.
These corrections are reflected in the
values presented in Table~\ref{abundtab}.

Finally, we adopt corrections for 
deviations from local thermodynamic equilibrium (LTE)
for a few of the light elements examined:\
Li \citep{lind09},
O \citep{fabbian09},
Na \citep{lind11},
Al \citep{andrievsky08},
and 
K \citep{takeda02}.
These corrections are also reflected in the abundances
presented in Table~\ref{abundtab}.
We apply no non-LTE corrections for heavy-element abundances.
Non-LTE corrections have been derived for individual lines of
a few heavier elements in other stars, but
published grids---like those available for the
light elements noted here---do not yet exist
to the best of our knowledge.

\section{Results}
\label{results}

\subsection{Possible Trends with Wavelength}
\label{fewave}

Offsets of varying magnitude
between the abundances of lines derived from optical
and NUV spectra have been noted in recent years
\citep{roederer10a,roederer12d,roederer14d,roederer14f,
lawler13,wood13,wood14,placco14,placco15,sneden16}.
These offsets are frequently found by comparing the
abundance derived from many Fe~\textsc{i} lines at 
a wide range of wavelengths.
The causes of these offsets are not understood at present,
but missing sources of continuous opacity and 
departures from LTE
may each be partly responsible.

Our initial list of Fe~\textsc{i} and \textsc{ii} lines in the NUV
was much longer than the final list presented in Table~\ref{fetab}.
Lines with obvious blends have been discarded,
and lines that yielded abundances deviant by more than
2$\sigma$ from the mean abundance were 
iteratively culled from the list.
Our final list contains 53 Fe~\textsc{i} lines and 40 Fe~\textsc{ii}
lines with wavelengths between 2070~\AA\ and 3100~\AA.~
The list of optical Fe~\textsc{i} and \textsc{ii} lines
(96 and 10~lines, respectively)
is taken from \citet{roederer14c} and 
rederived using the model atmosphere described in Section~\ref{atm}.

\begin{figure}
\begin{center}
\includegraphics[angle=0,width=3.3in]{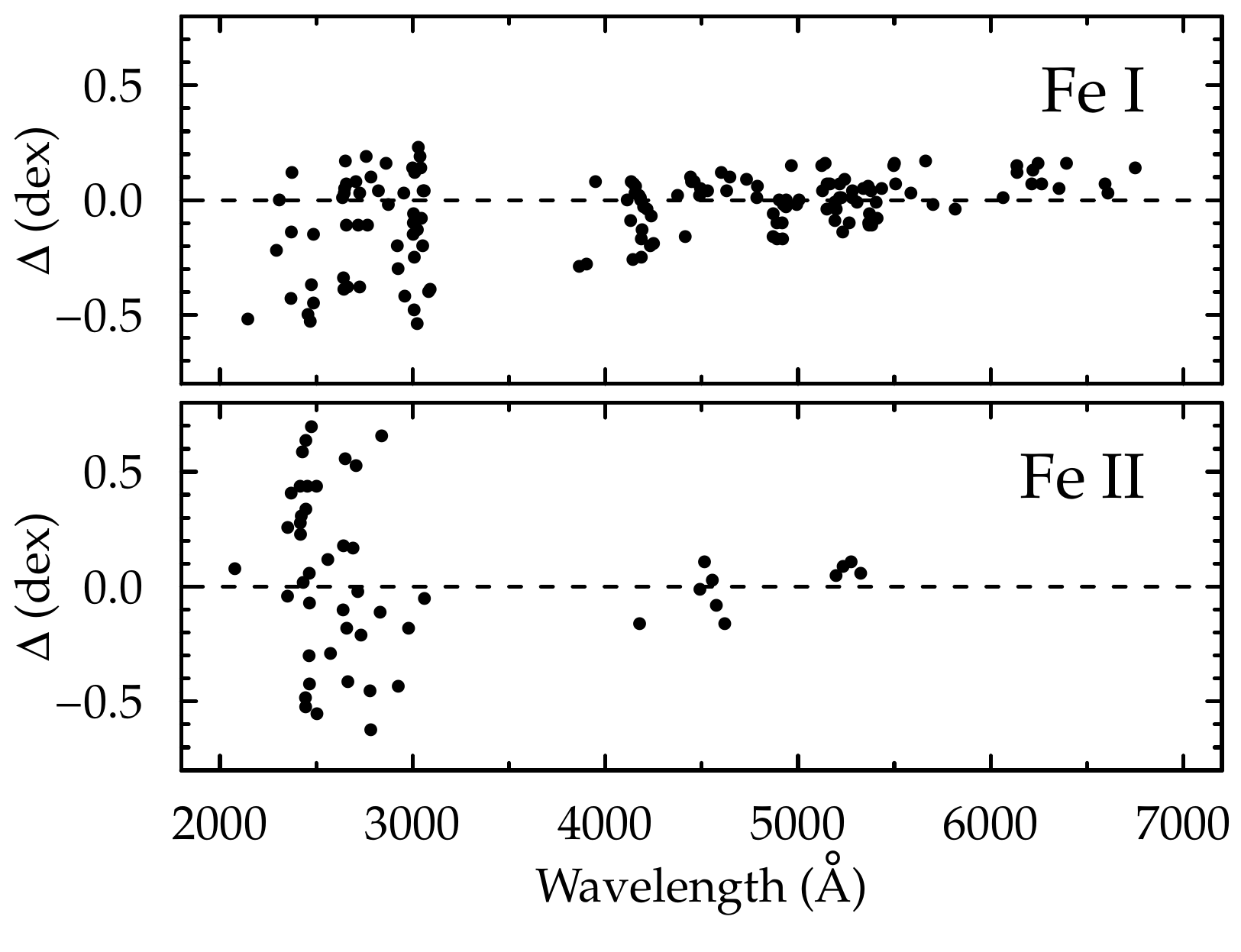}
\end{center}
\caption{
\label{feplot}
Fe abundance trends, relative to the mean abundance
for transitions with $\lambda >$~3800~\AA\ (dashed line),
as a function of wavelength.
 }
\end{figure}

Figure~\ref{feplot} illustrates our results.
The mean abundance derived from the NUV Fe~\textsc{i} lines is
[Fe/H]~$= -$1.94~$\pm$~0.03 ($\sigma =$~0.23), 
which is moderately lower than that derived from
optical Fe~\textsc{i} lines,
[Fe/H]~$= -$1.81~$\pm$~0.01 ($\sigma =$~0.11).
The mean abundance derived from the NUV Fe~\textsc{ii} lines is
[Fe/H]~$= -$1.52~$\pm$~0.06 ($\sigma =$~0.38),
which is in agreement with that derived from
optical Fe~\textsc{ii} lines,
[Fe/H]~$= -$1.57~$\pm$~0.03 ($\sigma =$~0.10).
Note that the scatter is considerably larger among the NUV lines,
which probably reflects unidentified blends
and the challenge of identifying the continuum in the NUV.~

If missing continuous opacity is the cause of the small Fe~\textsc{i}
abundance offset, we would expect a similar offset for all 
abundances, including 
those derived from
Fe~\textsc{ii} lines.
No offset is observed in the case of 
abundances derived from Fe~\textsc{ii} lines.
If non-LTE or another effect specific to Fe~\textsc{i} 
is the cause of the 
discrepancy, 
there is no need to apply a
global correction
to the abundances derived from lines of other species.
Lacking compelling evidence,
we do not
apply any corrections to the
abundances derived from lines in the NUV.~

\subsection{Heavy Elements in the NUV}
\label{newuv}

\citet{roederer12c} derived abundances of Ge, As, Se,
and Pt in \hd.
We expand this inventory to include
Cu, As (including one more line), 
Mo, Ru, Pd, Ag (upper limit only),
Cd, Te, Yb, Lu, W (upper limit only),
Os, Pt (one more line),
Hg (upper limit only), and Pb
(one more line).
These abundances are listed in Table~\ref{abundtab}.
The [X/Fe] ratios, where X represents a given element,
are computed using the Solar reference abundances
given in \citet{asplund09}.
All [X/Fe] ratios are 
constructed using the abundances derived from 
species in the same ionization state;
i.e., neutrals to neutrals and ions to ions.
Abundances or ratios denoted with the ionization state
are defined to be 
the total elemental abundance derived from transitions of
that ionization state 
after ionization corrections have been applied.

Table~\ref{abundtab} also lists several sets of uncertainties.
The statistical uncertainty, $\sigma_{\rm stat}$, 
is computed from equation~A17 of \citet{mcwilliam95},
which includes uncertainties in the equivalent widths or
synthesis matching
and \loggf\ values.
The total uncertainty, $\sigma_{\rm tot}$, 
is computed from equation~A16 of \citeauthor{mcwilliam95},
which includes the statistical uncertainty and 
uncertainties in the model atmosphere parameters.
The other two uncertainties listed in Table~\ref{abundtab}
are computed from approximations of 
equations~A19 and A20 of \citeauthor{mcwilliam95},
which give uncertainties in the abundance ratios.
To calculate the uncertainty in 
the ratio of two elements A and B, $\sigma_{\rm [A/B]}$,
we recommend that $\sigma_{\rm I}$ for element A 
be added in quadrature with $\sigma_{\rm stat}$ for
element B when element B is
derived from lines of neutral species.
Similarly,
we recommend that $\sigma_{\rm II}$ for element A 
be added in quadrature with $\sigma_{\rm stat}$ for
element B when element B is
derived from lines of ionized species.

\begin{figure}
\begin{center}
\includegraphics[angle=0,width=3.3in]{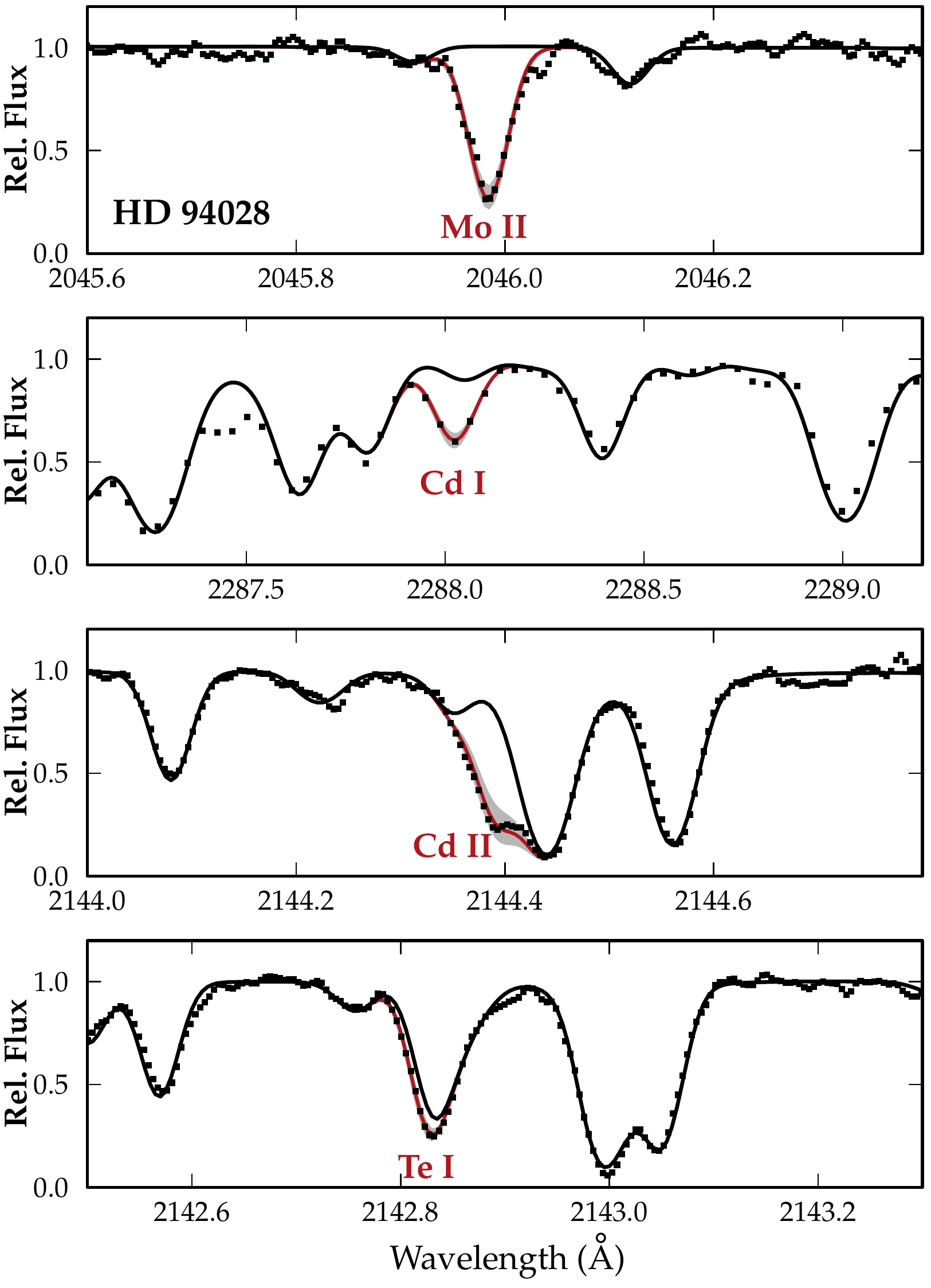}
\end{center}
\caption{
\label{specplot1}
Comparison of observed and synthetic spectra 
for a selection of Mo~\textsc{i}, Cd~\textsc{i},
Cd~\textsc{ii}, and Te~\textsc{i} lines.
The squares mark the observed spectrum of \mbox{HD~94028}.
The red line marks the best-fit abundance,
and the gray shaded region represents
a change in the best-fit abundance 
by a factor of $\approx$~two ($\pm$~0.3~dex).
The bold black line represents a synthesis with
no Mo, Cd, or Te present.
In the lower panel, the gray shaded region around the
Te~\textsc{i} line is largely hidden by the
red line,
which indicates that this line has little abundance sensitivity
as noted in the text.
 }
\end{figure}

The Te~\textsc{i} line at 2142.822~\AA\ is extremely
blended with an Fe~\textsc{i} line at 2142.832~\AA, 
as shown in Figure~\ref{specplot1}.
The \loggf\ value of the blending Fe~\textsc{i} line is 
unconstrained by experiment.
We follow our earlier work on \hdonesix\ \citep{roederer12b} 
and attempt to fit the overall line profile using the
abundance of Te and the \loggf\ value of the Fe~\textsc{i} line
as free parameters in the fit.
The \loggf\ value favored by our best fit,
$-$2.7, is moderately different than the value favored by our 
best fit in \hdonesix, $-$3.1. 
We attribute only a small fraction of the absorption at this wavelength
to Te~\textsc{i}, 
and the line has little sensitivity to the Te abundance,
thus 
we caution that the Te abundance derived from this line
is highly uncertain.

\begin{figure}
\begin{center}
\includegraphics[angle=0,width=3.3in]{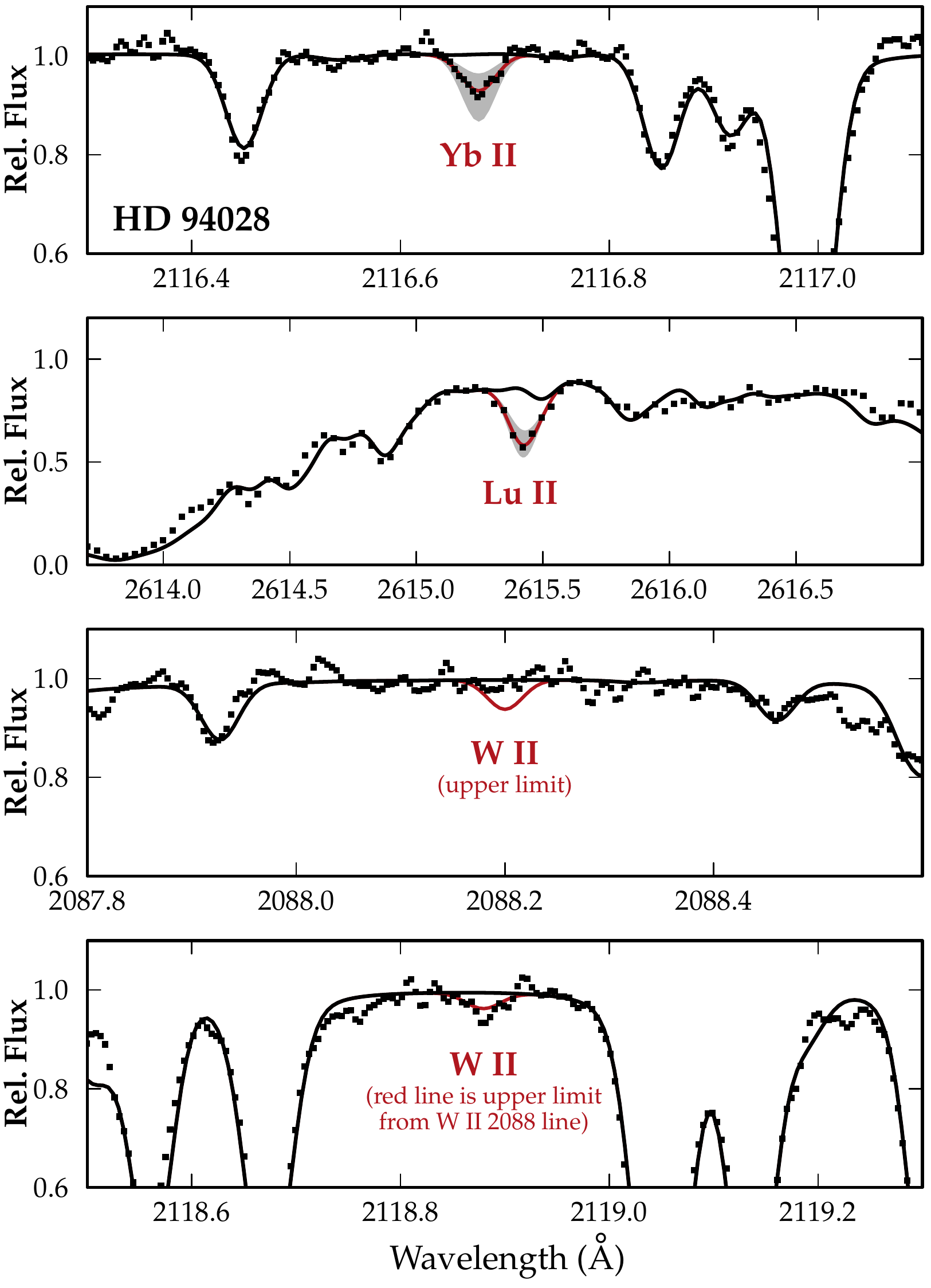}
\end{center}
\caption{
\label{specplot2}
Comparison of observed and synthetic spectra 
for a selection of Yb~\textsc{ii}, 
Lu~\textsc{ii}, and W~\textsc{ii} lines.
The squares mark the observed spectrum of \mbox{HD~94028}.
For Yb~\textsc{ii} and Lu~\textsc{ii},
the red line marks the best-fit abundance,
and the gray shaded region represents
a change in the best-fit abundance 
by a factor of $\approx$~two ($\pm$~0.3~dex).
For W~\textsc{ii}, the red line represents the
upper limit derived from the line at 2088.20~\AA.
The bold black line represents a synthesis with
no Yb, Lu, or W present.
 }
\end{figure}

The abundance derived from the Lu~\textsc{ii} line at 2615.41~\AA\
is higher than would be expected based on the
abundances of neighboring elements Yb and Hf
(Section~\ref{agbnucleo}).
We have no reason to discount our Lu measurement,
shown in Figure~\ref{specplot2},
but we caution that it is derived from a single line.
Other observations of the Lu~\textsc{ii} line at 2615~\AA\
would be useful to diagnose any unidentified systematics.

Figure~\ref{specplot2} illustrates the two lines of W~\textsc{ii}
that we have examined.
We do not detect absorption coincident with the W~\textsc{ii} line
at 2088.02~\AA,
and we derive an upper limit from this line.
We do, however, detect absorption coincident with the W~\textsc{ii}
line at 2118.88~\AA.
The upper limit derived from 
the former line
is inconsistent with the absorption detected at 
the latter.
The red line shown in the bottom panel of Figure~\ref{specplot2}
is the 3$\sigma$ upper limit inferred from the line
in the third panel of Figure~\ref{specplot2},
not a fit to the line shown.
We conclude that the absorption detected at 2118.88~\AA\ 
is not due to W~\textsc{ii},
so we derive only an upper limit on the W abundance.

\begin{figure}
\begin{center}
\includegraphics[angle=0,width=3.3in]{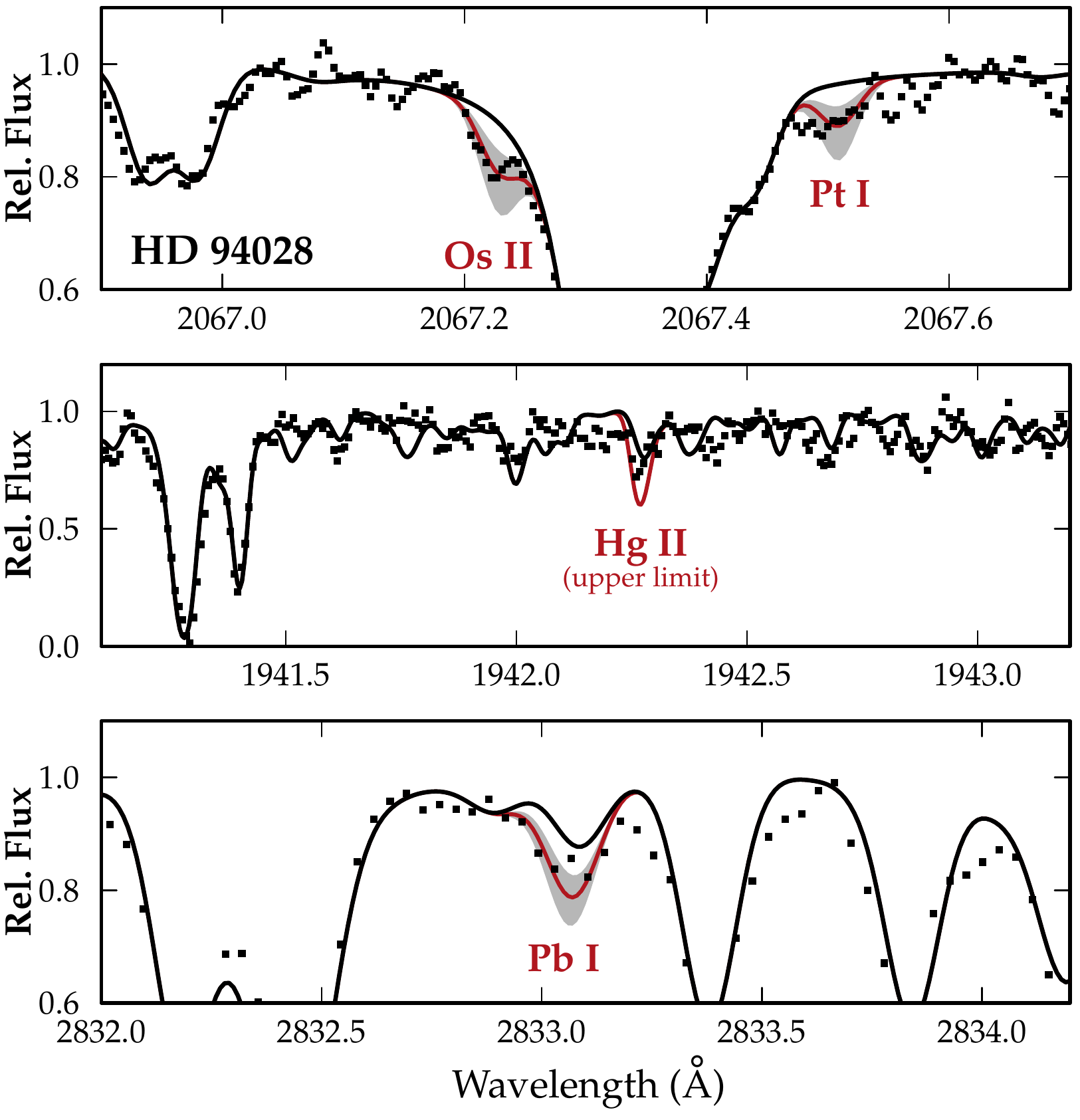}
\end{center}
\caption{
\label{specplot3}
Comparison of observed and synthetic spectra 
for a selection of Os~\textsc{ii}, Pt~\textsc{i},
Hg~\textsc{ii}, and Pb~\textsc{i} lines.
The squares mark the observed spectrum of \mbox{HD~94028}.
For Os~\textsc{ii}, Pt~\textsc{i}, and Pb~\textsc{i},
the red line marks the best-fit abundance,
and the gray shaded region represents
a change in the best-fit abundance 
by a factor of $\approx$~two ($\pm$~0.3~dex).
for Hg~\textsc{ii}, the red line represents
the upper limit derived assuming Hg~\textsc{ii} is
responsible for all absorption
at 1942.27~\AA.~
The bold black line represents a synthesis with
no Os, Pt, Hg, or Pb present.
 }
\end{figure}

The Hg~\textsc{ii} line at 1942.27~\AA\ presents an interesting case.
Absorption is clearly detected at this wavelength,
but as shown in Figure~\ref{specplot3}, 
our syntheses provide a poor fit to the spectral region 
surrounding this line.  
The S/N here is low ($\sim$~10--15).
Other hints of absorption at 1941.61~\AA,
1941.84~\AA, 1941.96~\AA, 1942.43~\AA, 
1942.66~\AA, and 1942.90~\AA\
correspond to $^{12}$C$^{16}$O transitions.
Fe~\textsc{ii} may also contribute to the 
absorption at 1942.99~\AA.
These absorption features also appear in a
spectrum of \hdonesix\ of similar quality,
so they are likely to be real lines and not noise.
Another $^{12}$C$^{16}$O transition 
of comparable strength
is predicted at 1942.27~\AA,
which coincides with the Hg~\textsc{ii} line.
Our syntheses fail to adequately reproduce the
strengths and the wavelengths of these CO lines.
The absorption at 1942.27~\AA\ is slightly stronger 
than the other predicted CO lines,
so Hg~\textsc{ii} might be present here;
however, we advise against attempts to 
derive an abundance from this blended, noisy line.
The upper limit on Hg that we report in Table~\ref{linetab}
is based on the entire absorption feature at 
1942.27~\AA\ resulting from 
an \spro\ isotopic mixture of Hg~\textsc{ii}.

\subsection{Multiple Species of Cu, Zn, Mo, and Cd}
\label{moly}

With access to the NUV spectrum of \hd, 
we have now detected both the neutral and singly-ionized states of 
Cu, Zn, Mo, and Cd.
Both states give abundances in agreement for Cu and Zn.
[Mo~\textsc{ii}/Fe] is higher than 
[Mo~\textsc{i}/Fe] by 
0.42~$\pm$~0.28~dex,
and 
[Cd~\textsc{ii}/Fe] is lower than
[Cd~\textsc{i}/Fe] by
0.59~$\pm$~0.33~dex.
\citet{roederer14d} derived [Cd/Fe] ratios
from Cd~\textsc{i} and \textsc{ii} lines
in two stars, and in both cases they agreed.

\citet{peterson11} derived abundances 
in \hd\ using the same STIS spectra we have used.
All abundance ratios in common with \citeauthor{peterson11}
are in excellent agreement.
The surprising result of her study was
an enhanced [Mo/Fe] ratio in \hd, $+$1.0, 
which we confirm,
$+$0.97~$\pm$~0.16.
Both results are derived from Mo~\textsc{ii} lines in the NUV.~
When one optical Mo~\textsc{i} line is used,
the [Mo/Fe] ratio is lower by a little
more than 1$\sigma$,
[Mo/Fe]~$= +$0.55~$\pm$~0.22.
\citet{roederer12b} and 
\citet{roederer14d} derived [Mo/Fe] ratios
from neutral and ionized lines in two stars,
and in all cases the [Mo/Fe] ratios
derived from different species agreed.

We suggest that the differences for [Cd/Fe] and [Mo/Fe] 
reflect the difficult nature of deriving
abundances from small numbers of lines
in crowded spectral regions.
Ideally, these ratios should be 
examined in larger samples of stars 
to identify potential systematic effects.
In the subsequent discussion,
we adopt the abundances derived from neutral lines for Zn,
ionized lines for Cu and Mo, 
and the average of the two for Cd.

\section{Discussion}
\label{discussion}

\subsection{AGB Models}
\label{models}

Substantial portions of
some of the heaviest elements---like 
Ba, Ce, and Pb---may be
produced by the $s$~process in
a star (or stars) that passed through the
thermally-pulsing AGB (TP-AGB) phase of evolution.
The $s$~process occurs in the 
He-intershell, 
where freshly synthesized elements are brought to the
surface by recurrent mixing episodes. 
Convection also dredges up
lighter elements such as C and F, 
which are produced together
with the $s$~process
\citep[see reviews by][]{herwig05,karakas14dawes}.

We use the low-metallicity AGB evolution and nucleosynthesis models from 
\citet{karakas14} and \citet{shingles15}. 
These models were calculated with an 
initial metallicity of $Z=0.0006$ ([Fe/H]~$=-1.4$), 
which is close to the derived 
metallicity of \hd.
The models cover a range in mass from 1.7\msun,
with a lifetime of 1.4~Gyr, to 6\msun, with a lifetime of 61~Myr. 
The low-mass models ($M \lesssim$~3\msun) 
require a \iso{13}C pocket for
the formation of \spro\ elements, 
and the predicted distribution at 
these low metallicities is peaked at Ba and Pb \citep{busso01,travaglio04}. 
In contrast, the He-intershells of the 
intermediate-mass models ($M >$~3\msun)
are hot enough to ignite the \iso{22}Ne($\alpha$,$n$)\iso{25}Mg reaction
more efficiently.
This typically produces elements at the first 
\spro\ peak near Rb and Sr.

Many uncertainties affect the AGB model predictions, 
including mass loss and convection.  
The AGB phase is terminated by mass loss, which also therefore 
determines the maximum enrichment. 
AGB stars in low-metallicity environments 
show similar mass-loss rates to AGB stars
in our Galaxy \citep{sloan09,lagadec09}. 
We use the \citet{vw93} mass-loss rate,
which has an implicit metallicity dependence in that
the equation for the pulsation
period depends on the radius. 
More thermal pulses are predicted for the lower
metallicity models \citep[e.g.,][]{karakas14b} 
compared with models at solar metallicity.
This may indicate that our mass-loss rates may be too low for most of the
AGB phase.  
Reducing the AGB lifetime would decrease the number of 
mixing episodes and would 
lower the yields of C and \spro\ elements.
The treatment of convection and non-convective mixing processes 
is also uncertain. 
We refer to \citet{karakas14dawes} for a detailed discussion, 
but we note
that including extra mixing in the envelopes of precursor
low-mass AGB stars may lower the final yield of C.~

\subsection{Comparison with AGB Nucleosynthesis Predictions}
\label{agbnucleo}

\begin{figure}
\begin{center}
\includegraphics[angle=0,width=3.3in]{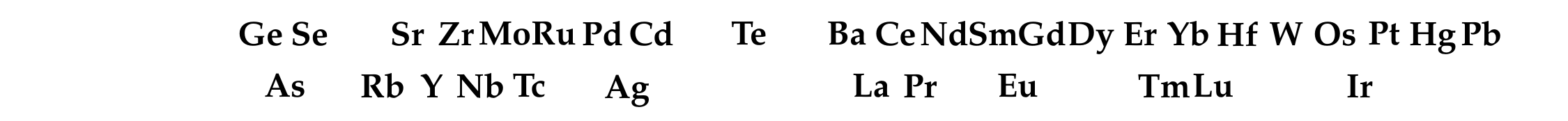} \\
\vspace*{-0.1in}
\includegraphics[angle=0,width=3.3in]{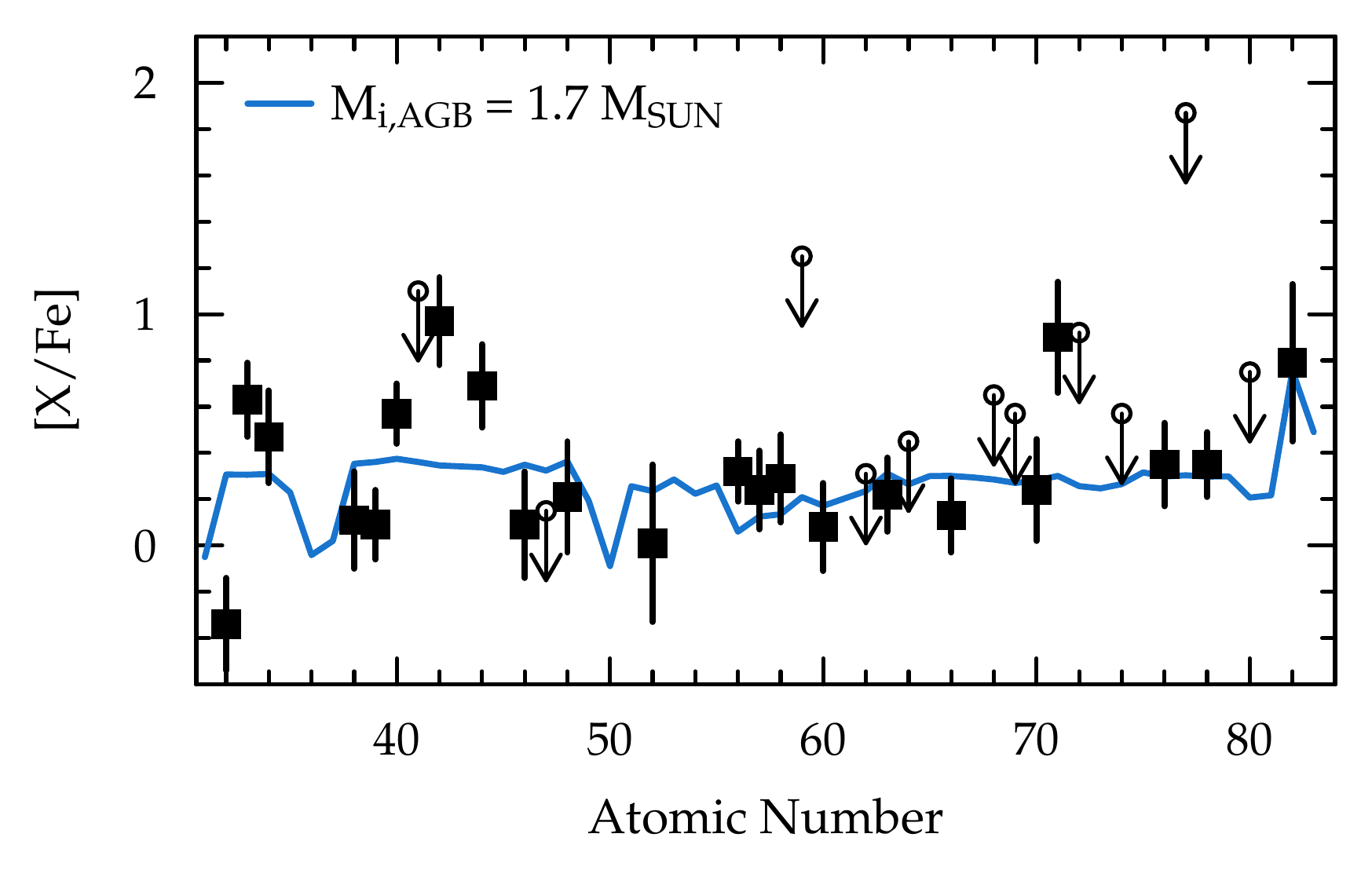} \\
\vspace*{-0.1in}
\includegraphics[angle=0,width=3.3in]{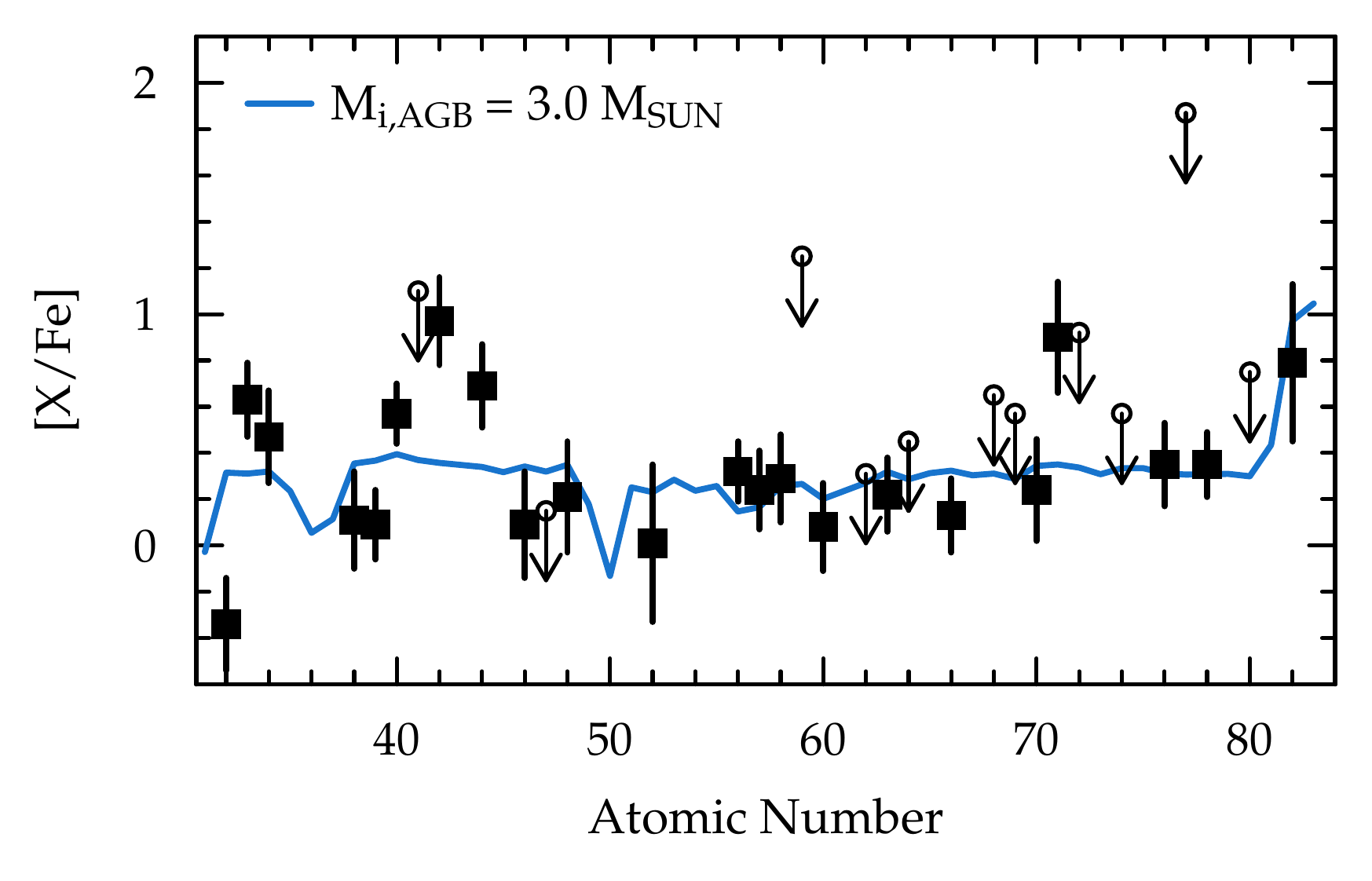} \\
\vspace*{-0.1in}
\includegraphics[angle=0,width=3.3in]{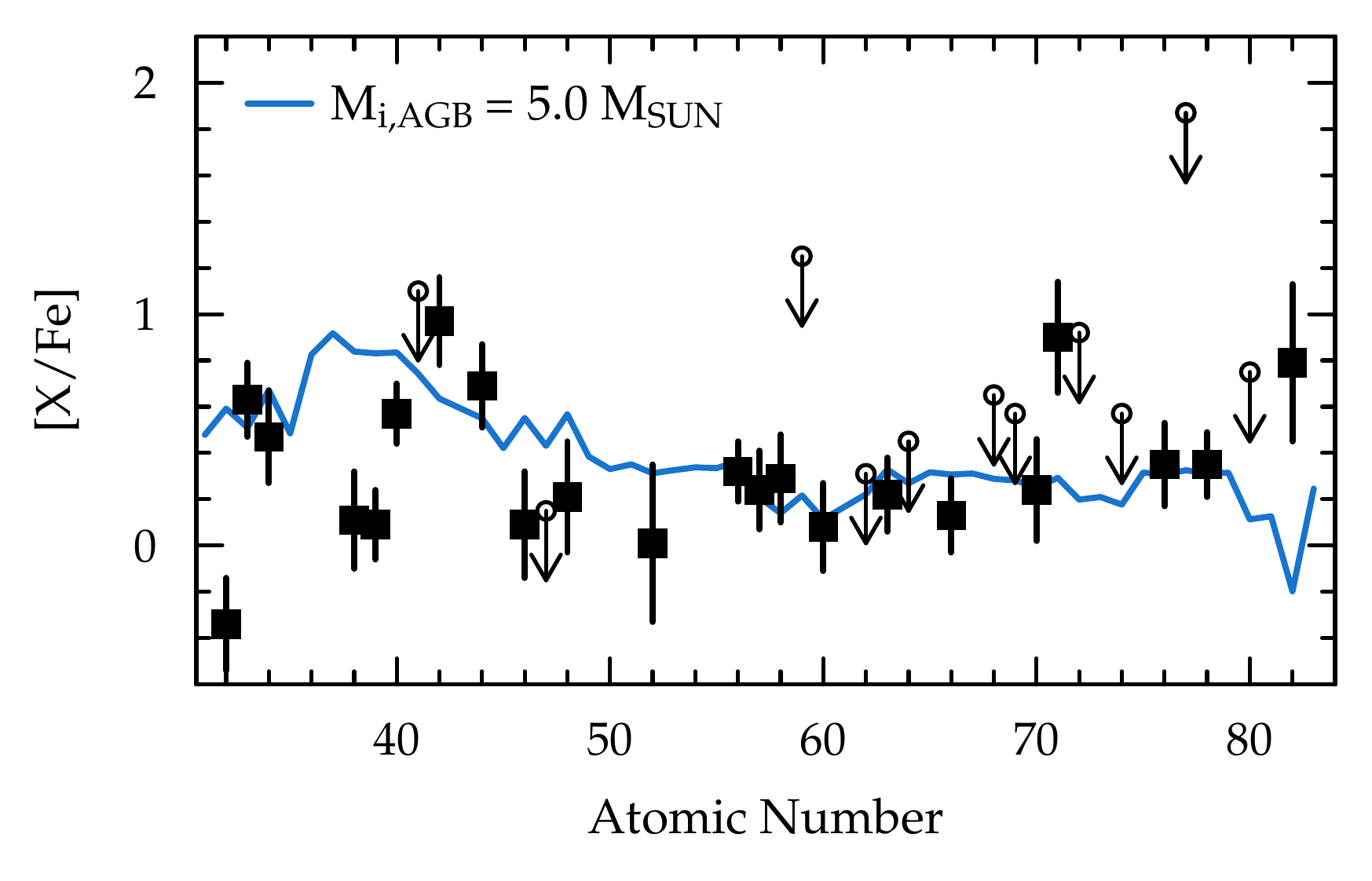} \\
\vspace*{-0.1in}
\includegraphics[angle=0,width=3.3in]{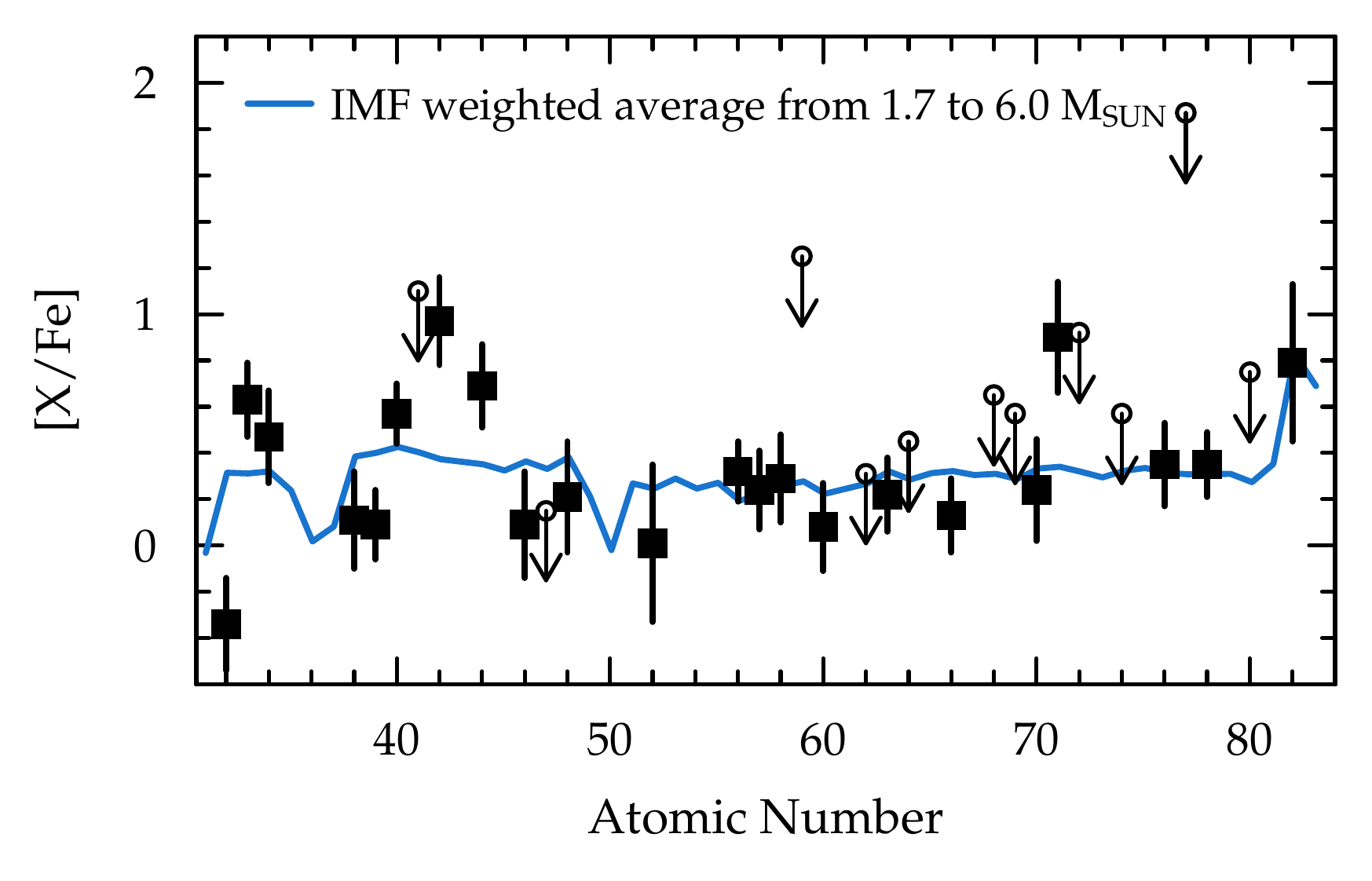} 
\end{center}
\caption{
\label{heavymodelplot}
Comparison of derived abundances (black symbols) and four
model fits (lines).
The final ejected AGB \spro\ yields have been diluted by 1.2\%,
1.2\%, 13\%, and 1.8\% for the 1.7~\msun, 
3.0~\msun, 5.0~\msun, and IMF-weighted models.
An \rpro\ component, scaled up by a factor of 2
relative to \mbox{HD~108317}, is included in each fit.
 }
\end{figure}

We illustrate the heavy element abundance pattern
in \hd\ in Figure~\ref{heavymodelplot}.
Most heavy elements, X, are only moderately overabundant,
with 0~$<$~[X/Fe]~$<$~1.
We compare this pattern with six 
sets of low-metallicity ([Fe/H]~$= -$1.4)
AGB nucleosynthesis yields, with $Y =$~0.24,
taken from \citet{karakas14} and \citet{shingles15}.
Several example fits are shown.
To create the fits shown in Figure~\ref{heavymodelplot}, 
we mix the final ejected AGB yields
with an \rpro\ ``foundation'' of material
(e.g., \citealt{bisterzo12,lugaro12}).
Nearly all metal-poor stars contain
some \rpro\ material (e.g., \citealt{roederer13}),
and we assume that \hd\ also inherited
\rpro\ material from its natal cloud.
We refer to this \rpro\ material as the \rpro\ foundation.

The 
pattern of the
\rpro\ abundance foundation is
inferred by
a two step process.
We begin by adopting 
the solar \rpro\ residual pattern.
This pattern is suspect
for elements below the second \rpro\ peak,
so we modify it for
32~$\leq Z \leq$~34 and 38~$\leq Z \leq$~48.
The modifications are made
using abundances from the metal-poor \rpro-enhanced star \hda\
\citep{roederer12d,roederer14d}.
In other words, each modified element X 
is normalized to Eu in the \rpro\ residual pattern
using the X/Eu ratio in \hda.

We adjust the relative proportions of the \rpro\ foundation
and the \spro\ material
by eye
to provide reasonable fits to the observed abundance pattern.
These proportions are well constrained by 
the elements beyond the second \rpro\ peak,
many of which 
are predominantly produced by only one of the two processes.
Elements whose production is dominated by
the $r$~process (e.g., Eu, Dy, Pt)
are reasonably fit by scaling up the \rpro\
foundation by a factor of two relative to \hda.
Dilution factors for the \spro\ material range from
1.2\% for the 1.7~\msun\ model to 16\% for the 6.0~\msun\ model.
In other words, only a small percentage of the \spro\ material
ejected by the AGB models is incorporated into \hd.
We cannot reproduce the overall shape of the abundance pattern
using any combination of \rpro\ and \spro\
curves for the 4.0, 5.0, and 6.0~\msun\ AGB models.
As shown for the 5.0~\msun\ model, the first \spro\
peak is over-produced by 0.2 to 0.8~dex, and the 
Pb abundance is under-produced by 1.0~dex.
The lower-mass AGB models provide better fits for most elements.

A fourth panel in Figure~\ref{heavymodelplot}
illustrates a scenario in which
\hd\ is assumed to have acquired its \spro\ material
from its natal cloud.
In this case, the \spro\ contribution to the model curve
shown reflects the yields of AGB stars from 1.7 to 6.0~\msun,
weighted by their relative numbers in a 
\citet{salpeter55} IMF
and the ejected Ba mass.
The dilution of the integrated \spro\ yields 
in this case is 1.8\%.
This also provides a reasonable fit for most
abundances, which is not surprising since
the low-mass AGB stars dominate by number and 
total mass of \spro\ material ejected per star.

\subsection{Possible Nucleosynthesis Scenarios}
\label{problems}

Our Ge, As, Mo, Ru, Ag, and Lu predictions
deviate from the $s$- and \rpro\ fits
by more than 1.5~times
the observational uncertainties.
No model or combination of models can simultaneously 
reproduce the enhanced [Zr/Fe], [Mo/Fe], and [Ru/Fe] ratios
and the solar [Sr/Fe], [Y/Fe], [Pd/Fe], [Ag/Fe], and [Cd/Fe] ratios.
The super-solar [As/Ge] ratio, 
$+$0.99~$\pm$~0.23~dex,
is also incompatible with 
\spro\ nucleosynthesis
or any reasonable combination of 
\spro\ and \rpro\ material.
The use of the un-modified \rpro\ residuals 
does not improve the fit for these elements,
particularly in the Sr--Ru mass region.
In this section, we investigate and exclude
several nucleosynthesis processes
as the sources of these ratios.

The [As/Ge]~$= +$0.99 and [Se/As]~$= -$0.16 
ratios effectively exclude 
several possible scenarios.
These are not compatible with the classical weak $r$~process
and different types of neutron-rich neutrino wind 
components in core-collapse supernovae (CCSNe;
e.g., \citealt{frohlich06,farouqi09,roberts10,arcones11}).
This signature is also not found in earlier 
electron-capture SN simulations 
(e.g., \citealt{hoffman08,wanajo09}).
More recent two-dimensional hydrodynamic 
explosion models of electron-capture SNe 
also produce [As/Ge]~$\ll$~1 \citep{wanajo11a}. 
Models for nucleosynthesis in proton-rich 
neutrino wind components predict a large scatter
among the relative ratios of Ge, As, and Se,
and the integrated [(Ge$+$As$+$Se)/Fe] ratio in
CCSN ejecta is not often available
(e.g., \citealt{kratz08,roberts10,arcones11,wanajo11b,arcones12}).
Previously, most studies have focused on the Sr--Ag mass region
(e.g., \citealt{hansen12}).
Now that observational data are available for metal-poor stars,
we recommend 
checking whether possible combinations
of neutrino-driven wind components could
explain the observed ratios
in the Ge--Se mass region.
Another scenario that could be explored
is the neutrino-induced $r$~process
in the He-shell of CCSNe,
where neutrons are made by the interaction
between neutrinos and He nuclei \citep{banerjee11}.

\begin{figure}
\begin{center}
\includegraphics[angle=0,width=3.3in]{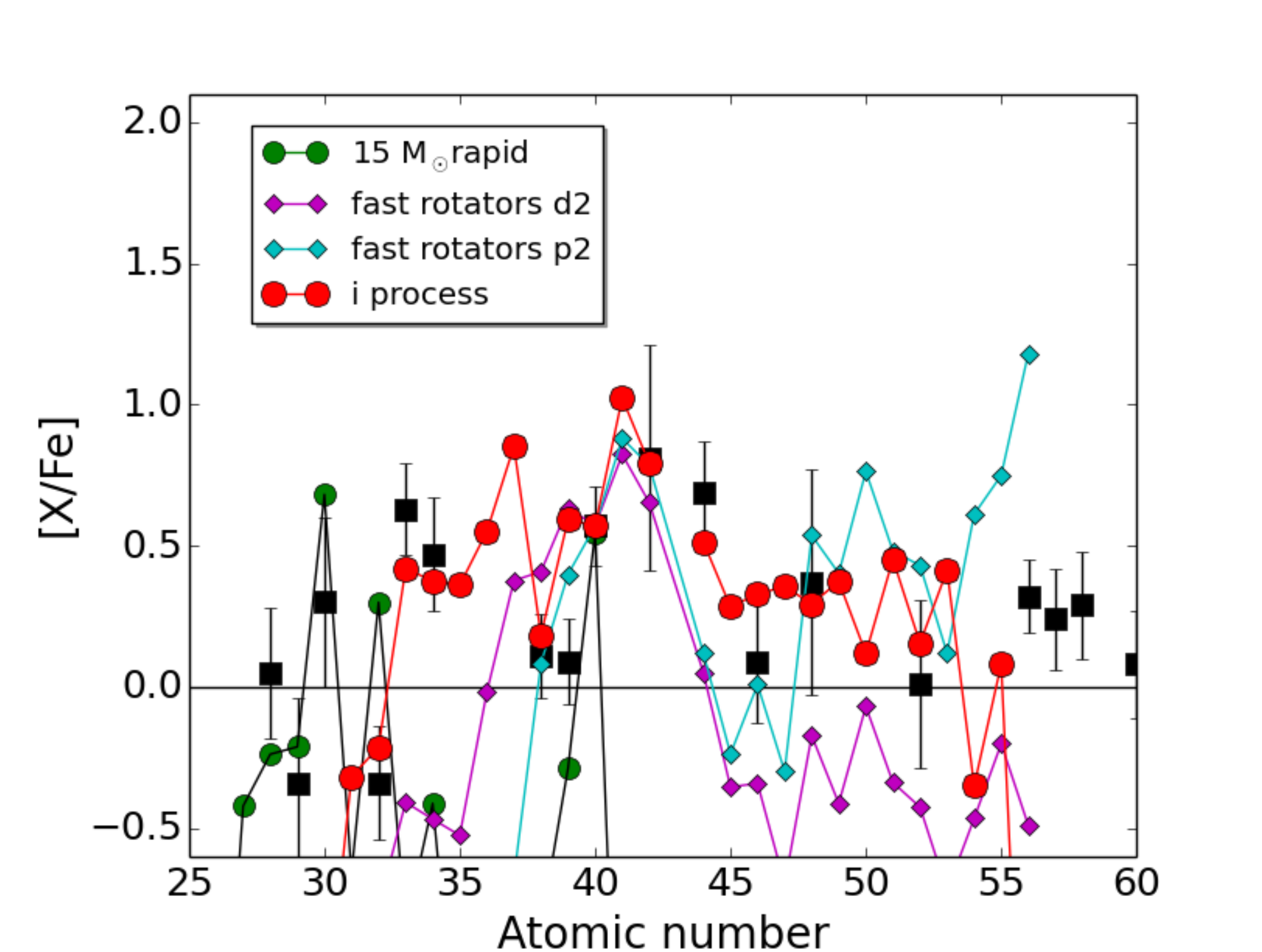}
\end{center}
\caption{
\label{marcoplot}
Comparison of the observed abundance pattern in \mbox{HD~94028}
with several model predictions
for the mass region from Co to Ba.
The observed abundance ratios are marked by black squares.
The small green circles show the predictions for yields from
the 15~\msun\ CCSN model \citep{pignatari13},
the purple and teal diamonds show the 
predictions for yields from the fast-rotating massive stars
\citep{pignatari08} multiplied (``p2'') and 
divided (``d2'') by a factor of 2, 
and the large red circles show the predictions 
for yields from the $i$~process.
Black squares mark the observed abundance ratios.
The solid line marks the solar ratios.
}
\end{figure}

The $\alpha$-rich freezeout component 
is made in the deepest ejecta of CCSNe
(e.g., \citealt{woosley92}).
It is thought to be 
the main source of $^{64}$Zn in the solar system
and Zn in the early Galaxy
(e.g., \citealt{nomoto13}).
In Figure~\ref{marcoplot}, 
we compare the abundances in \hd\ with the products of a 
15~\msun\ CCSN model by \citet{pignatari13},
which includes $\alpha$-rich freezeout ejecta.
The progenitor structure was computed using
the stellar code 
GENEC \citep{eggenberger08}, and the CCSN explosion simulations 
include the fallback prescription ``rapid'' according to 
\citet{fryer12}.
The predictions are inconsistent with the observations.
The $\alpha$-rich freezeout
component in these models mainly produces Zn, Ge, and Zr, 
and the production of As is negligible. 
These yields are affected by many uncertainties, 
including the details of the explosion,
fallback, and progenitor mass.
The ability to predict the relative production
of elements in different mass regions
(e.g., Zn and Zr) is limited.
Predictions for neighboring elements produced in similar conditions
should be more reliable, and
the enhanced [As/Ge] ratio
found in \hd\ is 
inconsistent with theoretical predictions for the
$\alpha$-rich freezeout \citep{woosley92}.

The AGB yields discussed in Section~\ref{agbnucleo}
all predict $-$0.2~$\leq$~[As/Ge]~$\leq$~0.0.
The low [As/Ge] predictions are mainly due to the 
local \ncap\ cross sections in the 
Ge--As mass region along the \spro\ path.
As is not made efficiently
due to the relatively high 
$^{75}$As($n$,$\gamma$)$^{76}$As
cross section
(e.g., \citealt{dillmann08,marganiec09}).
This conclusion is independent of the stellar source
of the $s$~process or
stellar uncertainties affecting the 
\spro\ production at low metallicity 
(e.g., \citealt{gallino98,straniero06,bisterzo12,lugaro12,karakas14dawes}).
In other words, 
the nuclear physics 
limits the amount of As
that the $s$~process can make.  

The $s$~process
in fast-rotating massive stars 
has been proposed as one of the possible scenarios 
to explain a sample of old, metal-poor stars
with enhanced Sr, Y, and Zr 
(e.g., \citealt{pignatari08,frischknecht12}).
The boosted $s$~process at low metallicity 
is due to the primary production of 
$^{14}$N and $^{22}$Ne, and $^{22}$Ne is
the main neutron source in these stellar models.
In Figure~\ref{marcoplot}, we compare the abundances
in \hd\ with two models of fast-rotating massive stars by 
\citet{pignatari08}, using the 
$^{22}$Ne($\alpha$,$n$)$^{25}$Mg rate
multiplied and divided by a factor of 2 
(see discussion in \citealt{pignatari08} for details).
The abundance pattern 
is not compatible with the observations in \hd.
In particular, the models cannot reproduce the [Mo/Ru] ratio, 
and they do not efficiently produce elements in the Ge--As region. 
The [As/Ge] ratio is sub-solar for both models,
as expected for an \spro\ source.

\subsection{The $i$-process Contribution}
\label{iprocess}

A scenario that may be compatible with the 
observed abundance pattern among the lighter \ncap\
elements in \hd\ 
is the $i$~process \citep{cowan77}.
The $i$~process is a \ncap\ process
triggered by the rapid ingestion of 
a substantial quantity of H in He-burning convective regions. 
H is captured by $^{12}$C to produce $^{13}$N. 
The $^{13}$N decays to $^{13}$C, 
and the 
$^{13}$C($\alpha$,$n$)$^{16}$O
reaction is efficiently activated.
The conditions leading to \ipro\ nucleosynthesis
may be found in multiple stellar sites,
including super-AGB and post-AGB stars,
He-core and He-shell flashes in 
low-metallicity low-mass stars, and
massive stars.

The H-ingestion episodes that activate the $i$~process
are quite different from those that form the 
$^{13}$C pockets in AGB stars.
$^{13}$C pockets form when
a small amount of H is mixed 
into the He-burning region on timescales of 
$\sim$~10$^{2}$~yr, 
comparable to the timescale for dredge-up episodes,
which lead to neutron densities 
$\sim 10^{8}$~cm$^{-3}$.
The key for the $i$~process is that the neutron density
is governed by the interplay of two 
coincidentally similar timescales 
(see \citealt{herwig11} for more details). 
The first is the convective
turn-over timescale of $\sim$~10--20~min,
where the ingested H is advected 
into deeper and hotter layers of the He-burning convection zone. 
$^{13}$N is released when
the timescale of the $^{12}$C($p$,$\gamma$)$^{13}$N reaction
equals this convective mixing timescale.
The second is the $\beta^{+}$ decay timescale of $^{13}$N, 
9.6~min,
which limits how fast $^{13}$C can be produced.
Generally speaking,
the robustness of the \ipro\ neutron density 
is based on the universal nature of these timescales, 
which ensure that a neutron density $\sim 10^{15}$~cm$^{-3}$
is produced 
whenever H is ingested into He-burning driven convection.

In contrast, the neutron exposure does change from 
one \ipro\ site to another, since
the \ipro\ conditions are inherently three-dimensional and 
hydrodynamic in nature. 
The $^{12}$C($p$,$\gamma$)$^{13}$N 
reaction releases
large amounts of energy 
that affect the convective flow. 
Present models \citep{herwig14} 
indicate that the star responds with a 
global oscillation of shell H-ingestion. 
Its non-spherical nature 
implies that the stellar response 
to the energy release from H-ingestions 
cannot be known from spherically-symmetric simulations alone. 
Hydrodynamic simulations 
indicate that the stellar response 
may be violent, 
and it is possible that the hydrodynamic feedback 
will terminate the conditions suitable for the $i$~process.
This termination would be different for each site
(e.g., low-mass AGB stars, post-AGB stars, or super-AGB stars). 
The result would be nucleosynthetic signatures 
that share a common intermediate neutron density 
but differ substantially in neutron exposure.
For example, 
we observe overabundances at and beyond Ba 
in some \mbox{CEMP-$r$/$s$} stars,
suggestive of an $i$~process with a large neutron exposure
\citep{dardelet15}.
We observe a significant enhancement of first-peak elements 
in post-AGB stars, 
suggestive of an $i$~process with a lower neutron exposure
\citep{herwig11}.

These three-dimensional hydrodynamical models 
can be used to inform one-dimensional stellar models
\citep{herwig11,stancliffe11,herwig14,woodward15}
and calculations involving the \ipro\ products
(e.g., \citealt{cristallo09,herwig11,bertolli13}).
\citeauthor{bertolli13}\ provided an \ipro\ trajectory
to explore the main properties of this process.
This one-zone trajectory has, by design, 
the same neutron density that we associate 
with convective ingestion of H into a He-burning layer. 
We use this trajectory
and assume a stellar metallicity of [Fe/H]~$= -$2
to calculate
a set of \ipro\ abundance ratios.
Figure~\ref{marcoplot} shows 
the results of this calculation.
We adjust the conditions (neutron exposure and 
termination time) to 
maximize production in the As--Mo region.
These conditions also produce super-solar [As/Ge] ratios,
and production drops beyond Te.
The \ipro\ fit is not perfect.
For example, it over-produces Y compared to the observations.
This is a simple trajectory, 
and more comprehensive models will be necessary
to verify or refute this scenario. 
Nevertheless, 
the $i$~process
provides a scenario that can simultaneously 
explain the [As/Ge] and [Se/As] ratios
and the enhancement in Mo and Ru.

\subsection{Light Elements}
\label{light}

The elements with $Z \leq$~30 in \hd\
are all normal when compared with 
other main sequence or subgiant stars 
with similar metallicity.
This material 
presumably originated 
in prior generations of CCSNe.

The AGB models discussed in Section~\ref{agbnucleo}
also indicate that significant amounts of 
light elements (C to P, 6~$\leq Z \leq$~15)
are produced.
The light and heavy elements
produced by AGB stars should have the 
same dilution factor
when material is transferred to \hd\
or its natal cloud.
The dilution factors derived in the fits to the 
heavy elements are so small ($<$~2\%) 
that any contributions from the AGB stars
to N, O, Na, Mg, Al, Si, and P
are negligible ($\delta$[X/Fe]~$\leq$~0.03~dex) when compared with
the material produced in CCSNe.

The [C/Fe] ratio, $-$0.06~$\pm$~0.19, 
tells a different story.
This ratio is in agreement with
the mean [C/Fe] ratio for main sequence and subgiant
stars with [Fe/H]~$> -$2, [C/Fe]~$\approx -$0.12
(e.g., \citealt{gratton00,roederer14c}).
Our low-mass and IMF-weighted 
models over-predict the [C/Fe] ratio
for \hd\ by $\sim$~0.4~dex
after dilution.
In other words, a self-consistent enrichment scenario
for the C and \spro\ material in \hd\
would predict that \hd\ is mildly C-enhanced,
whereas no C enhancement is observed.

This discrepancy is not resolved by 
shortcomings of our model atmosphere.
Comparisons of [C/Fe] ratios derived from CH lines in
one- and three-dimensional
model atmosphere calculations of metal-poor dwarfs
reveal large discrepancies
\citep{asplund04,behara10}.
They indicate that one-dimensional models
overestimate the [C/Fe] ratio, however,
so this is unlikely to be the source of the discrepancy.

Our [C/Fe] ratio is in agreement with previous
analyses of CH molecular bands in \hd.
\citet{gratton00} derived [C/Fe]~$= -$0.07~$\pm$~0.09,
\citet{simmerer04} derived [C/Fe]~$= +$0.05~$\pm$~0.2, and
\citet{lai07} derived [C/Fe]~$= -$0.05~$\pm$~0.17.
\citet{tomkin92} examined both CH and C~\textsc{i} features
in \hd, finding that [C/Fe]~$= -$0.2~$\pm$~0.15 from the molecular lines,
$+$0.32~$\pm$~0.09 from the atomic lines (assuming LTE),
and 
$+$0.28~$\pm$~0.09 from the atomic lines (assuming non-LTE).
They found offsets
between the CH and C~\textsc{i} abundance indicators
in other stars in their sample, but
the results from the atomic lines
showed a dependence on \teff.
\citeauthor{tomkin92}\ suggested that the abundances derived from
the C~\textsc{i} lines were too high and affected by a systematic error.
We conclude that [C/Fe] is not enhanced above the solar ratio
in \hd,
so the tension with our AGB model predictions stands.

The degree of the C discrepancy could be reduced
if extra mixing occurs in the
envelopes of low-mass AGB stars, 
as discussed in Section~\ref{models}.
Alternatively,
the discrepancy could be resolved if 
no AGB stars contributed to \hd.
This seems unlikely, however, 
since the elements at the second and third \spro\ 
peaks are enhanced.

\subsection{The Enrichment of HD~94028}
\label{enrichment}

\begin{figure}
\begin{center}
\includegraphics[angle=0,width=3.3in]{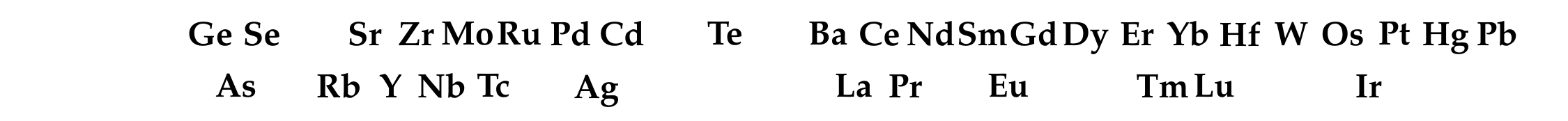} \\
\vspace*{-0.1in}
\includegraphics[angle=0,width=3.3in]{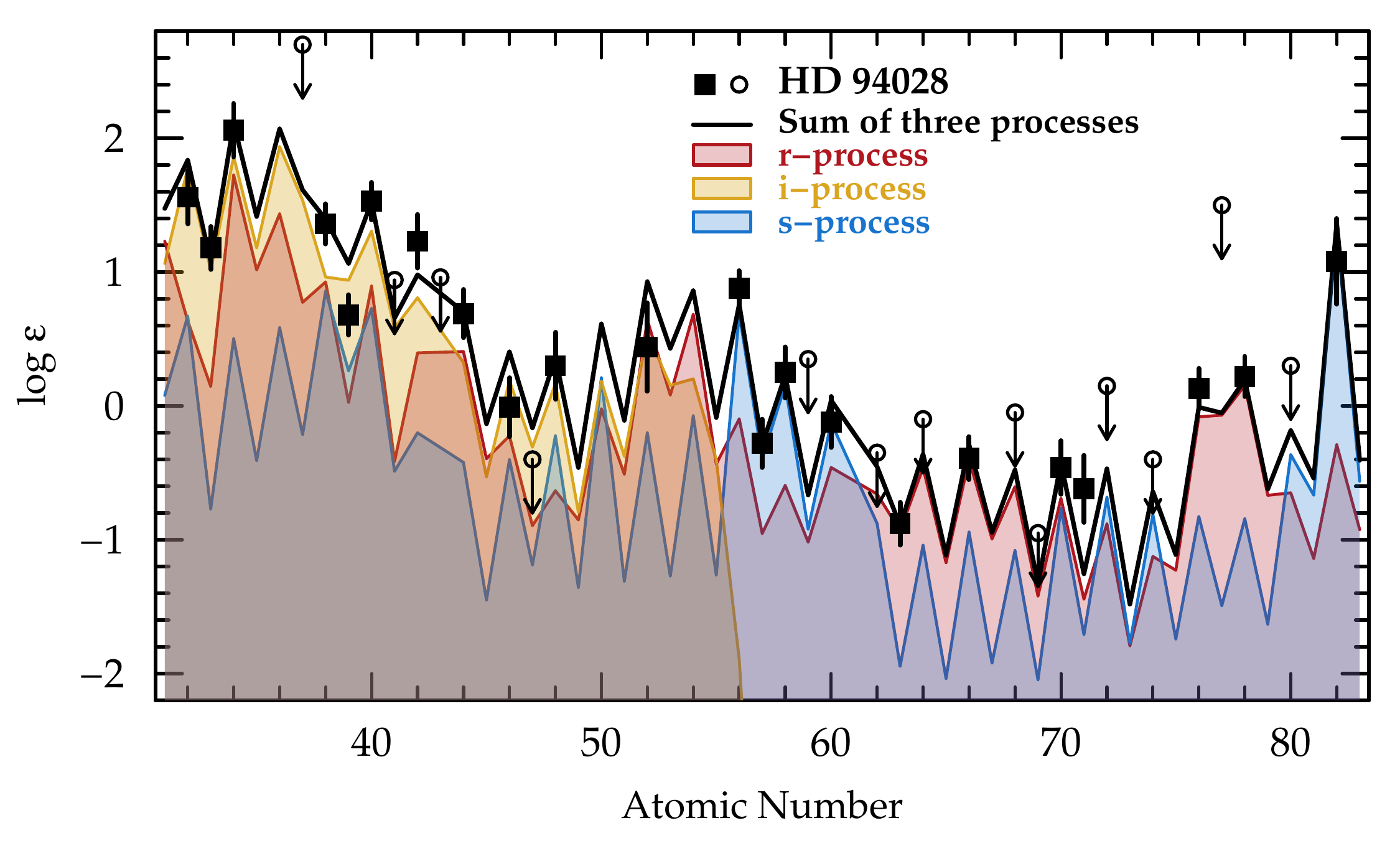} \\
\vspace*{-0.1in}
\includegraphics[angle=0,width=3.3in]{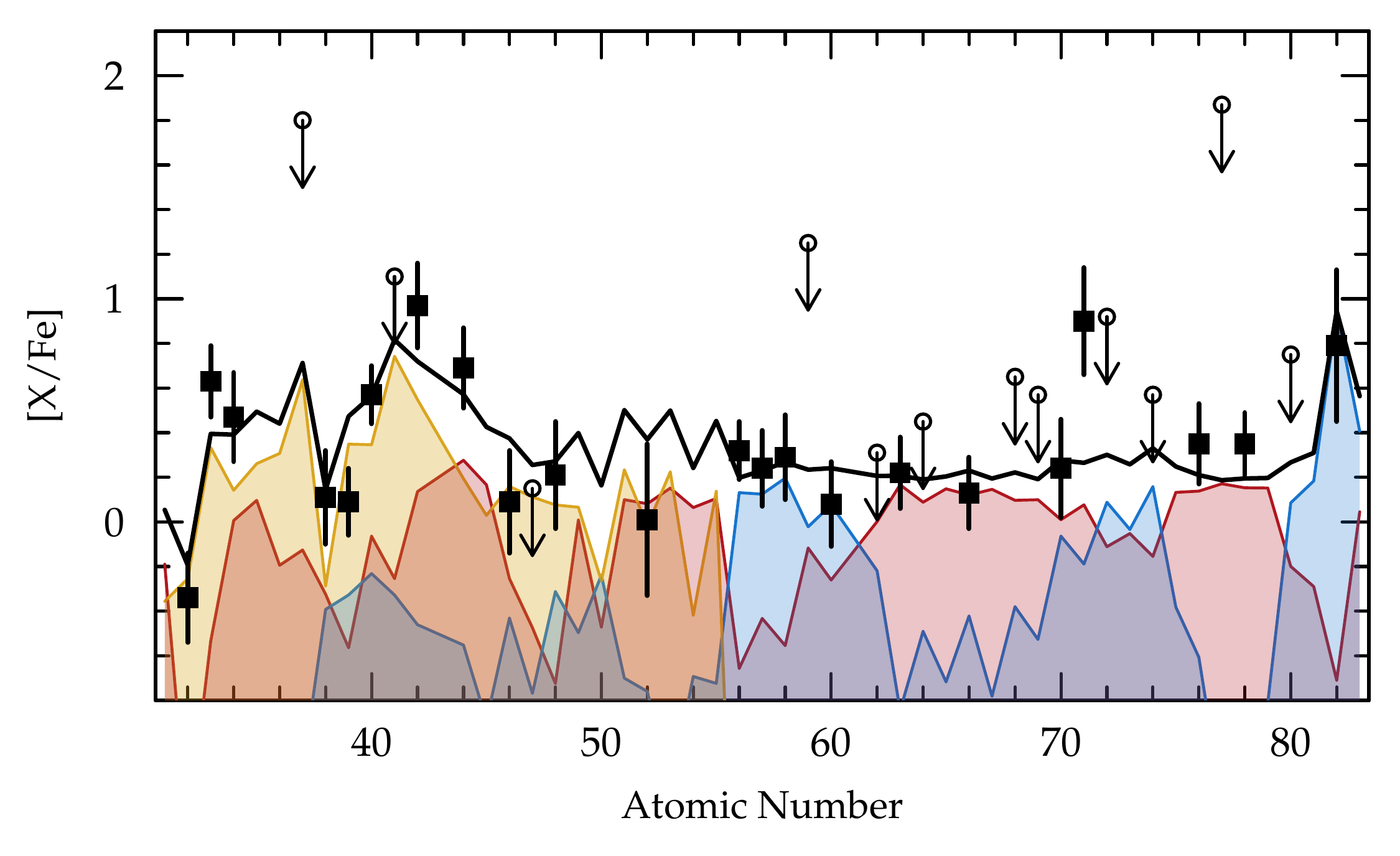} \\
\vspace*{-0.1in}
\includegraphics[angle=0,width=3.3in]{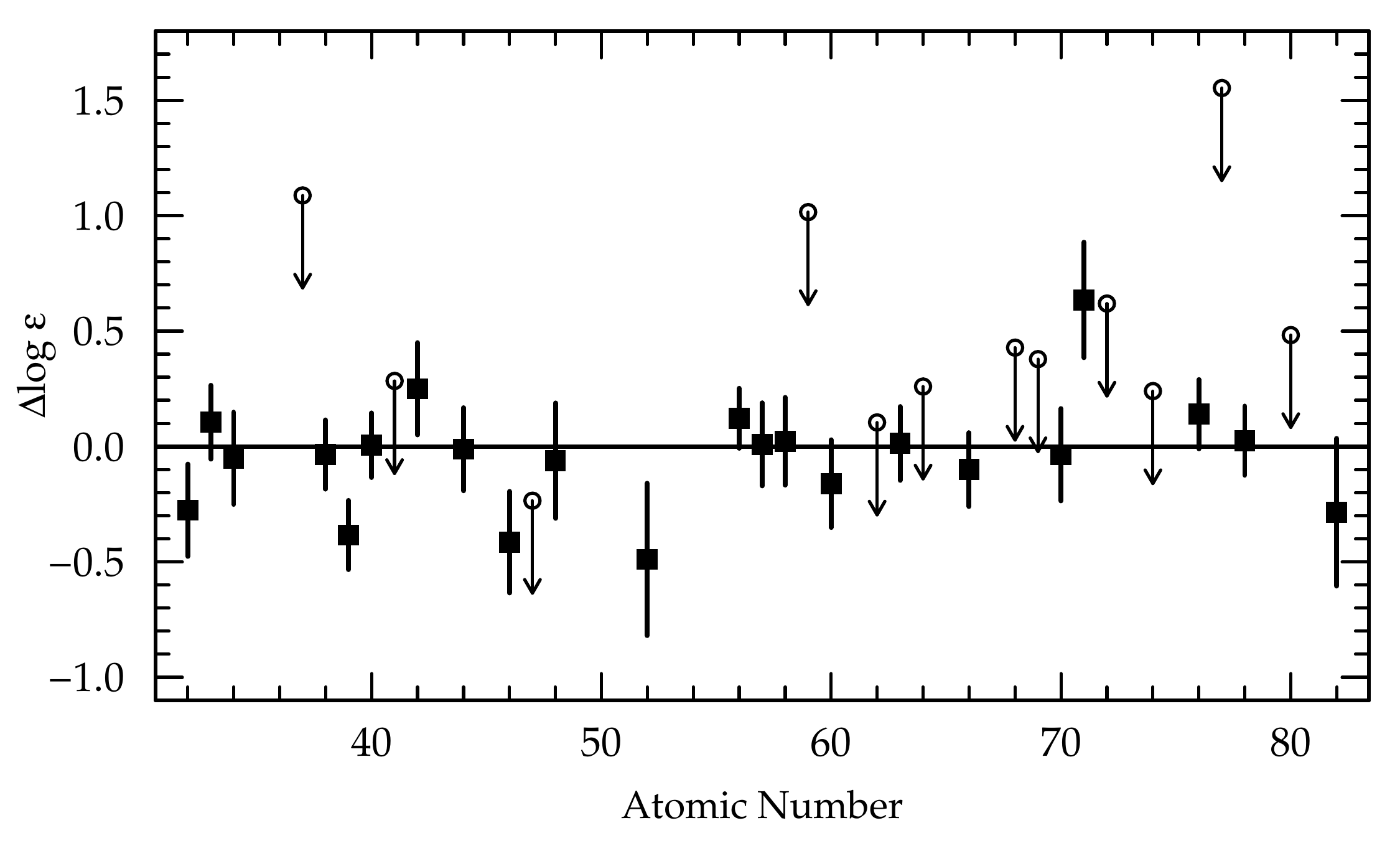} 
\end{center}
\caption{
\label{finalplot}
Comparison of the observed abundance pattern in \mbox{HD~94028}
and contributions from the 
\spro\ (blue),
\rpro\ (red), and
\ipro\ (gold) models.
The \spro\ component is taken from the 1.7~\msun\ TP-AGB model
discussed in Section~\ref{agbnucleo}.
The \rpro\ component
is based on the solar \rpro\ residuals and 
modified as described in Section~\ref{agbnucleo}.
The \ipro\ component is based on the trajectory 
from \citet{bertolli13}
and tuned to maximize production in the As--Mo region
as described in Section~\ref{iprocess}.
The solid black line marks the sum of the three
processes for each element.
The overall normalization for each process
has been adjusted by-eye.
The distribution is expressed as $\log\epsilon$ and [X/Fe]
in the top and middle panels,
and the bottom panel illustrates the residuals between the
observed abundance pattern and the sum of the models.
}
\end{figure}

In this section, we explore scenarios where the
\ncap\ abundances in \hd\ result from the
combined contributions of the 
$s$, $r$, and $i$~processes.
These are illustrated in Figure~\ref{finalplot}.
The agreement is superb for most elements,
and only Y, Pd, and Lu
deviate by $\sim$~2$\sigma$.
The $i$~process contributes the largest share
for most elements with $Z <$~50 in \hd.
An inferior fit is obtained if 
we omit any \rpro\ contribution to the elements
lighter than Te.
Using the un-modified \rpro\ residuals 
over-predicts the [Ge/Fe] and [As/Fe] ratios
but has no substantial impact on Se or the
Sr--Ru region.
The \ipro\ contribution
is still required
regardless of whether we use the modified or un-modified
\rpro\ pattern.
The heaviest elements ($Z \geq$~56)
owe their origin only to the
$r$~process and $s$~process.
The $r$~process dominates the production of elements 
near the rare-earth and third peaks,
while the $s$~process dominates the production
of the light rare-earth elements
and Pb.

Is \hd\ a single star, or does it have an unseen
white dwarf companion?
\hd\ shows no compelling evidence of radial velocity
variations;
measurements by \citet{latham02}
span more than 5100~d and have an rms
of only 0.64~\kmsec.
If, however, there is a companion
and the system is observed face-on, it would also
show no radial velocity variations.

Is the presence of \spro\ material evidence
that \hd\ must be in a binary system
with a white dwarf companion?
\citet{casagrande11} derived an age of \hd\ of
12.35~Gyr (7.5 to 13.8 Gyr at 95\% confidence intervals)
from comparison with Padova and BASTI isochrones.
The longest-lived AGB star considered in Sections~\ref{models}
and \ref{agbnucleo} has 
a lifetime of 1.4~Gyr.
Such stars could have formed, evolved through the TP-AGB
phase, and polluted the ISM before \hd\ was born.
In other words, the presence of \spro\ material
does not require a more-evolved companion star for \hd.

We are unable to exclude either
the single-star or binary system scenarios.
We expect that \hd\
acquired its \rpro\ material from its natal cloud.
The \spro\ and \ipro\ material may have also been
present in the natal cloud or 
added later by a companion.

\subsection{The $i$-process in Other Stars in the Early Galaxy}
\label{otherstars}

Evidence for the $i$~process has been observed in the
post-AGB star known as Sakurai's object 
(\object[V4334 Sgr]{V4334~Sgr}; \citealt{herwig11})
and in presolar grains found in pristine meteorites 
\citep{fujiya13,jadhav13,liu14}.
There are hints that
the $i$~process may also be responsible for some of the
abundance patterns observed
in young open clusters \citep{mishenina15},
the CEMP-$r/s$ stars \citep{dardelet15,jones15}, and
low-mass post-AGB stars in the Magellanic Clouds \citep{lugaro15}.

Our observations may be generalized to suggest that 
super-solar [As/Ge] and solar or sub-solar [Se/As] ratios
could signal the operation of
the $i$~process in the early Galaxy.
These are a common feature in the
nine metal-poor
stars analyzed by \citet{roederer12c} and \citet{roederer12d,roederer14d}.
The [As/Ge] ratios range from $+$0.65 to $+$1.00
in five stars with $-$2.5~$<$~[Fe/H]~$< -$0.5.
[As/Ge] is constrained to be $>+$0.75 in two other stars,
and only upper limits ($< +$1.34) are available for two more.
The [Se/As] ratios in these stars
range from $-$0.53 to $+$0.54 
with a mean of $-$0.10.
The observational uncertainties on these ratios are typically 
0.3--0.4~dex for [As/Ge] and 0.3--0.7~dex for [Se/As], which
reflects the challenge of measuring absorption lines
in the crowded regions of the NUV spectrum.

No non-LTE calculations exist for Ge~\textsc{i}, As~\textsc{i}, or 
Se~\textsc{i}
lines in late-type stars.
Ge has a lower first ionization potential (7.90~eV) than As (9.79~eV)
or Se (9.75~eV).
If overionization occurs, it is more
likely to preferentially affect Ge~\textsc{i} lines.
This would reduce the [As/Ge] ratios.
Non-LTE corrections for other species with 
low first ionization potentials 
are typically 0.1--0.2~dex and 
rarely exceed 0.5~dex 
(e.g., \citealt{takeda05,bergemann12,yan15})
in late-type stars, however.
We conclude that it is unlikely that non-LTE effects
can produce solar or sub-solar [As/Ge] ratios in these stars.

\citet{peterson11} pointed out that the 
enhanced [Mo/Fe] ratios found in \hd\ and \hdonesix\
were uncommon.
Subsequent data support this assertion.
Mo was detected in 30 of the 311 metal-poor
stars examined by \citet{roederer14d}.
Two of those 30 stars are highly enriched in \rpro\ material
(\object[BPS CS 22892-052]{CS~22892--052} and
 \object[BPS CS 31082-001]{CS~31082--001}),
and they are not representative of the
majority of metal-poor stars.
Among the remaining 28~stars, 
none show [Mo/Fe]~$>+$0.6.
Similarly, in the sample of stars with [Fe/H]~$< -$1
examined by \citet{hansen14}, 
only 4 of the 34 stars with detected Mo
show [Mo/Fe]~$>+$0.6.
In contrast, super-solar [As/Ge] and solar or sub-solar [Se/As] ratios
appear common, at least among the limited
metal-poor stars where Ge, As, and Se have been studied.
This contrast may illustrate the 
diverse paths leading to \ipro\ nucleosynthesis.

\section{Conclusions}
\label{conclusions}

We have performed a detailed
abundance analysis of elements
detectable in the NUV and optical spectra
(1885--8000~\AA) of the metal-poor main sequence star \hd.
We have derived abundances of many trace elements
that are rarely studied in late-type stars.
Our analysis reveals that several of these 
elements, including Ge, As, Se, and Mo,
hold important---and previously unrecognized---clues
necessary to identify the \ncap\ processes
responsible for creating the heavy elements
observed in \hd\ and other metal-poor stars.

We find that no combination of 
\rpro\ and \spro\ material can fully 
account for the abundances of elements
from Ge to Pb in \hd.
Some \rpro\ material is clearly present,
as revealed by the rare earth elements
and third \rpro\ peak.
The \spro\ pattern observed among other elements from
Ba to Pb favors relatively 
low-mass AGB stars ($\approx$~1.7~\msun\ or so),
although it is difficult to reconcile this
with the solar [C/Fe] ratio observed in \hd.
Including the contributions from another
\ncap\ process, the $i$~process,
improves the fit for elements from Ge to Te.
This is necessary to fit the
super-solar [As/Ge], 
solar or sub-solar [Se/As], 
and
enhanced [Mo/Fe] and [Ru/Fe] ratios.
We exclude scenarios involving 
the classical weak $r$~process,
$\alpha$-rich freezeout in CCSNe, and
the $s$~process in fast-rotating massive stars.
Other explosive-nucleosynthesis components 
in CCSNe and in electron-capture SNe 
need to be tested in detail for this mass region 
to verify which scenarios are compatible 
and which can be ruled out. 

Our analysis establishes a new constraint for stellar nucleosynthesis. 
These are the crucial questions:\
what are the nucleosynthesis conditions that lead to super-solar 
[As/Ge] and solar or sub-solar [Se/As] ratios,
and can some scenarios be excluded?
The Ge-As-Se mass region has never been a focus
of detailed analysis of the
different neutrino-wind nucleosynthesis components
because of a lack of observations.
The impact of internal stellar dynamics and 
nuclear uncertainties also need to be considered.
Here, our preliminary results indicate that the
$i$~process may be a good candidate to explain
these abundance patterns.

\acknowledgments

I.U.R.\ acknowledges generous support for program number AR-13879,
provided by NASA through a grant from the
Space Telescope Science Institute, which is 
operated by the Association of Universities for
Research in Astronomy, Incorporated, 
under NASA contract NAS5-26555.
I.U.R., M.P., and F.H.\ acknowledge partial support 
from grant PHY~14-30152 (Physics Frontier Center/JINA-CEE)
awarded by the U.S.\ National Science Foundation (NSF).~
A.I.K.\ was supported through an 
Australian Research Council Future Fellowship (FT110100475).
M.P.\ and F.H.\ 
acknowledge significant support to NuGrid from
NSF grants PHY~02-16783 and PHY~09-22648 (JINA)
and 
EU MIRG-CT-2006-046520. 
M.P.\ and F.H.\ also acknowledge 
the services of the Canadian Advanced Network 
for Astronomy Research (CANFAR), 
which in turn is supported by CANARIE, 
Compute Canada, University of Victoria, 
the National Research Council of Canada, and the Canadian Space Agency.
M.P.\ also acknowledges support from the ``Lend\"{u}let-2014'' 
Programme of the Hungarian Academy of Sciences (Hungary),
SNF (Switzerland), and the BRIDGCE UK network.
This research has made use of NASA's
Astrophysics Data System Bibliographic Services;
the arXiv pre-print server operated by Cornell University;
the SIMBAD and VizieR databases hosted by the
Strasbourg Astronomical Data Center;
the Atomic Spectra Database hosted by
the National Institute of Standards and Technology;
the Mikulski Archive for Space Telescopes
at the Space Telescope Science Institute;
IRAF software packages
distributed by the National Optical Astronomy Observatories,
which are operated by the Association of Universities for Research
in Astronomy, Inc., under cooperative agreement with the National
Science Foundation;
and the R software package \citep{r}.

{\it Facilities:} 
\facility{HST (STIS)},
\facility{Smith (Tull)},
\facility{VLT:Kueyen (UVES)}

\end{document}